%% file: main.tex
\documentclass[a4paper,fleqn]{cas-dc}
\usepackage{siunitx}

\usepackage{hyperref}
\usepackage[numbers, square, comma, sort&compress]{natbib}
\usepackage[normalem]{ulem} 
\usepackage{amsmath, amsthm}
\usepackage[english]{babel}
\newtheorem{theorem}{Theorem}

\usepackage{empheq}

\usepackage{xcolor}
\usepackage{ulem}
\usepackage{siunitx}
\usepackage{extdash}
\usepackage{comment}
\usepackage{soul}

\usepackage{bbm}
\usepackage{caption}
\usepackage{subcaption}
\usepackage{accents}
\newcommand{\ubar}[1]{\underaccent{\bar}{#1}}
\newcommand{\di}{n}
\newcommand{\tdi}{N}

\ifodd 1
    \newcommand{\arxiv}[1]{}
\else
    \newcommand{\arxiv}[1]{#1} 
\fi

\ifodd 1
    
    \newcommand{\comn}[1]{\textbf{\color{red} (COMMENT: #1)}} 
\else
    
    \newcommand{\comn}[1]{}
\fi

\ifodd 0
    \newcommand{\rev}[1]{{\color{blue}#1}}
    \newcommand{\com}[1]{\textbf{\color{red} (COMMENT: #1)}} 
\else
    \newcommand{\rev}[1]{{#1}}
    \newcommand{\com}[1]{}
\fi

\def\tsc#1{\csdef{#1}{\textsc{\lowercase{#1}}\xspace}}
\tsc{WGM}
\tsc{QE}
\tsc{EP}
\tsc{PMS}
\tsc{BEC}
\tsc{DE}
\usepackage{amsthm}

\newcommand{\be}{\begin{equation}}
\newcommand{\ee}{\end{equation}}
\newcommand{\bee}{\begin{eqnarray}}
\newcommand{\eee}{\end{eqnarray}}
\newcommand{\bse}{\begin{subequations}}
	\newcommand{\ese}{\end{subequations}}

\usepackage{bbm,epstopdf, algorithm, algpseudocode, multirow}

\newcommand{\remText}[1]{\iffalse {#1} \fi}

\newcommand{\Kh}{\si{\kelvin\hour}}

\newcommand{\rText}[1]{\iffalse {#1} \fi}

\usepackage{nomencl} 
\usepackage{framed} 
\usepackage{multicol} 
\makenomenclature
\renewcommand*\nompreamble{\begin{multicols}{2}}
\renewcommand*\nompostamble{\end{multicols}}

\begin{document}
\let\WriteBookmarks\relax
\def\floatpagepagefraction{1}
\def\textpagefraction{.001}
\shorttitle{Adaptive Constrained Tuning of Building Controllers}
\shortauthors{Wenjie Xu et~al.}


\title [mode = title]{Data-driven adaptive building thermal controller tuning with constraints: A primal-dual contextual Bayesian optimization approach}




\author[1,2]{Wenjie Xu}[orcid=0000-0001-7475-0056]
\credit{Conceptualization, Methodology, Software, Validation, Formal analysis, Data Curation, Visualization, Writing - Original Draft}

\author[1]{Bratislav Svetozarevic}[orcid=]
\credit{Conceptualization, Methodology, Writing - 
Review \& Editing, Supervision}

\author[1,2]{Loris Di Natale}
\credit{Software, Validation, Data Curation, Visualization, Writing - Review \& Editing}

\author[1]{Philipp Heer}
\credit{Writing - Review \& Editing, Resources, Funding acquisition}

\author[2]{Colin N Jones}
\credit{Conceptualization, Methodology, Writing - Review \& 
Editing, Supervision}

\address[1]{Urban Energy Systems Laboratory, Swiss Federal Laboratories for Materials Science and Technology (Empa), 8600 D\"{u}bendorf, Switzerland}
\address[2]{Laboratoire d'Automatique, Swiss Federal Institute of Technology Lausanne (EPFL), 1015 Lausanne, Switzerland}

\cortext[cor1]{Corresponding author:  \texttt{wenjie.xu@epfl.ch} (Wenjie Xu), Laboratoire d'Automatique, Swiss Federal Institute of Technology Lausanne (EPFL), 1015 Lausanne, Switzerland}



\arxiv{
\begin{graphicalabstract}
    \begin{figure*}[h]
    \begin{center}
    \includegraphics[width=\textwidth]{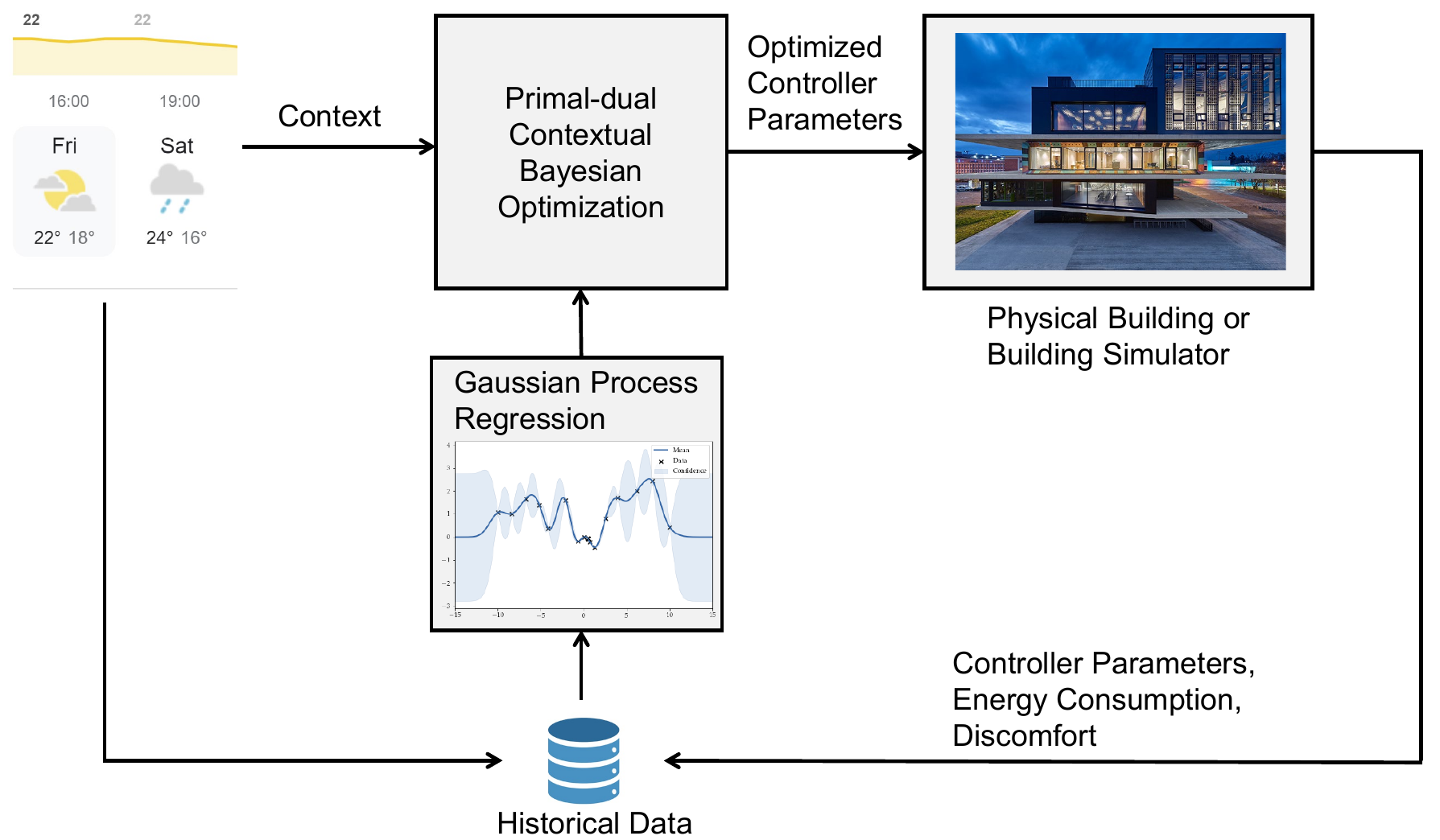}
    \label{fig: graphical abstract}
    \end{center}
    \end{figure*}

\end{graphicalabstract}


\begin{highlights}



{\item A primal-dual contextual Bayesian optimization framework \rev{to tune building controllers} \rev{is proposed}.}

{\item \rev{It a}\rev{symptotically {achieves the \emph{optimal} performance} while \emph{satisfying constraints} {on average}.} 
}



\item \rev{It saves up to} \rev{$4.7\%$} energy compared to alternatives\rev{, satisfying thermal comfort constraints on average.} 

\item \rev{Alternatively}\rev{, it can} reduce the \rev{thermal} discomfort by \rev{$63\%$} with\rev{in a given} energy budget. 

 


\end{highlights}
}



\begin{abstract}
We study the problem of tuning the parameters of a room temperature controller to minimize its energy consumption, subject to the constraint that the \rev{daily} cumulative \rev{thermal} discomfort \rev{of the occupants} 
\rev{is} below a given threshold. We formulate it as an online constrained black-box optimization problem where, on each day, we observe some relevant environmental context 
and adaptively select the controller parameters. 
{In this paper, w}e propose \rev{to use} a data-driven \textbf{P}rimal-\textbf{D}ual \textbf{C}ontextual \textbf{B}ayesian \textbf{O}ptimization~\rev{(PDCBO)} approach to solve this problem. 

In a {simulation} case study \rev{on a single room}, we apply our algorithm to tune \rev{the parameters of a Proportional Integral} (PI) \rev{heating} controller and the pre-heating time. Our results show that PDCBO can save up to $4.7\%$ energy consumption compared to \rev{other} state-of-the-art Bayesian optimization-based methods while keeping the {daily thermal} discomfort below the given tolerable threshold \rev{on average}. Additionally, \rev{PDCBO} 
can automatically track {time-varying} tolerable thresholds while existing methods fail to do so. \rev{We then} 
study an alternative 
constrained tuning problem where we aim to minimize the \rev{thermal} discomfort \rev{with a given energy budget}. With this formulation, PDCBO reduces the average discomfort by up to $63\%$ compared to 
state-of-the-art safe optimization methods while keeping the average \rev{daily} energy consumption below the \rev{required threshold.}
\end{abstract}


\maketitle

\input{introduction}
\makenomenclature

\begin{table*}[!t]   

\begin{framed}

\nomenclature[]{\textit{Symbols}}{}
\nomenclature[01]{$\di$}{Index of day} 
\nomenclature[02]{$E_\di$}{Energy consumption on day $\di$}
\nomenclature[03]{$D_\di$}{Discomfort on day $\di$}
\nomenclature[04]{$D^{\mathrm{thr}}$}{Discomfort threshold}
\nomenclature[05]{$T^{\textrm{ambient}}_\di$}{Forecasted average ambient temperature on day $\di$} 
\nomenclature[06]{$I_\di$}{Forecasted average solar irradiation on day $\di$}
\nomenclature[07]{$\theta$}{Controller parameters} 
\nomenclature[08]{$z$}{Contextual variables}
\nomenclature[09]{$J(\theta, z)$}{The energy consumption model function}
\nomenclature[10]{$g(\theta, z)$}{The discomfort model function}
\nomenclature[11]{$\nu^J_\di$}{Noise in the energy consumption on day $\di$}
\nomenclature[12]{$\nu^g_\di$}{Noise in the discomfort on day $\di$}
\nomenclature[13]{$\sigma_J$}{Standard deviation for $\nu^J$}
\nomenclature[14]{$\sigma_g$}{Standard deviation for $\nu^g$}
\nomenclature[15]{$\mathcal{H}_\di$}{$\left\{(E_\tau, D_\tau, T^{\textrm{ambient}}_\tau, I_\tau,T^{\textrm{init}}_\tau, \theta_\tau)_{\tau=1}^{\di-1}\right\}$}
\nomenclature[16]{$T^{\textrm{init}}_\di$}{Initial room temperature on day $\di$}
\nomenclature[17]{$t$}{Time in a day}
\nomenclature[18]{$u(t)$}{Normalized heating power at time $t$}
\nomenclature[19]{$T(t)$}{Zone temperature at time $t$}
\nomenclature[20]{$T_{\textrm{setpoint}}$}{Setpoint temperature}
\nomenclature[21]{$e(t)$}{Temperature error $T_{\textrm{setpoint}}-T(t)$ at time $t$}
\nomenclature[22]{$K_\textrm{p}$}{Proportional gain of PI controller}
\nomenclature[23]{$K_\textrm{i}$}{Integral gain of PI controller}
\nomenclature[24]{$T_\textrm{min}(t)$}{Comfort range lower bound at time $t$}
\nomenclature[25]{$T_\textrm{max}(t)$}{Comfort range upper bound at time $t$}
\nomenclature[26]{$\epsilon$}{Constant drift in the PDCBO algorithm}

\nomenclature[27]{\textit{Abbreviations}}{}
\nomenclature[28]{PDCBO}{Primal-dual Contextual Bayesian Optimization} 
\nomenclature[29]{GP}{Gaussian process}
\nomenclature[30]{PI}{Proportional-Integral}
\printnomenclature

\end{framed}

\end{table*}

\input{prob_statement}
\input{case_study}
\input{methodology}

\input{result}
\input{conclusion}

\input{acknowledgements}

\printcredits

\bibliographystyle{els-cas-templates/model1-num-names}

\bibliography{ref.bib}


\end{document}

%% file: introduction.tex
\section{Introduction}
With the increasing challenge of climate change, the high energy usage in the building sector calls for the development of \rev{energy-}efficient \rev{control} solutions. \rev{On the other hand, it is also critical to meet the comfort requirement of the occupants~\cite{boodi2018intelligent}.} 
%
%
While advanced control methods have been developed~(e.g., model predictive control~\cite{oldewurtel2012use,XIAO2023121165,GAO2023121106} and reinforcement learning~\cite{svetozarevic2022data,LEI2022119742,YANG2015577,CORACI2023120598,SHEN2022118724}) to optimize the energy consumption and\rev{/or the comfort of the occupants}, simple rule-based controllers~(e.g., bang-bang or Proportional Integral Derivative PID controllers) still dominate the current practice~\cite{svetozarevic2022data}. \rev{The parameters of these controllers are typically set manually by building engineers} \rev{and kept fixed for the life of the building}. However, these \rev{hand-tuned} simple rule-based controllers 
have {two main} limitations in practice.

     The first limitation is {\emph{sub-optimality}}. \rev{It is generally hard to explicitly model the effect of given control parameters on the performance (the energy consumption and thermal comfort of the occupants) \rev{to optimize them}. Consequently, manually tuned} controllers \rev{usually} operate suboptimally~\cite{salsbury2005survey, stluka2018architectures}, potentially incurring excessively high discomfort or energy consumption.
     
     The other limitation is {\emph{non-adaptivity}}. \rev{Given the influence of external factors, typically the weather conditions,} an optimal building control policy should intuitively \rev{be able to} adapt to environmental changes. Previous works indeed indicated that adaptive controller parameters can bring significant energy savings~(e.g., in \cite{fiducioso2019safe}). However, the parameters of rule-based controllers are typically set during commissioning and then kept fixed. 

\rev{Meanwhile, with more and more ubiquitous sensing technology~\cite{yan2023ai}, we can monitor the operations of buildings online. By collecting the temperature and heating power trajectories, building control performance, including energy consumption and occupant comfort, can be evaluated on each day.}

Based on \rev{the above observations}, it is of great practical interest to \rev{be able to} adaptively tune building controller parameters to \rev{optimize their online performance}. 
\rev{Since we are interested in minimizing both the energy consumption and the thermal discomfort of the occupants,} \rev{tuning these parameters is a multi-objective problem}. \rev{Furthermore, the two objectives are often in contradiction, with improved thermal comfort \rev{typically} coming at the cost of additional energy consumption, making the tuning procedure challenging in general.} 

\rev{In this work, we formulate \rev{it} 
as an online} constrained optimization \rev{problem}, where one metric \rev{--- the energy consumption or the thermal discomfort ---} is set as the objective and the other as the constraint. \rev{On each day, \rev{we solve an} online problem 
to get a new set of controller parameters and apply them to the building. 
} However, solving \rev{this constrained problem} poses several new challenges.
    
    \textbf{\emph{\rev{Constraint} violation management.}} 
    It is \rev{generally} difficult\rev{, if even possible,} to \rev{strictly satisfy the comfort or energy constraints} 
    at all times \rev{since the effect of the control parameters on the energy consumption or the thermal comfort is hard to predict accurately. 
    However, thanks to the soft nature of these constraints, some violations are acceptable \rev{in practice}. Interestingly,} 
    \rev{short-term} constraint violations \rev{can be beneficial} to \rev{efficiently} identify the optimal set of parameters, \rev{as shown in} previous works~\cite{xu2022vabo,xu2023violation}. Despite the benefits of short-term constraint violation, we still want to ensure constraint satisfaction \rev{in the long term}. 
    
    \textbf{\emph{Cumulative constraint satisfaction.}} 
   \rev{While the thermal comfort of the occupants has to be monitored every day, a}n energy consumption constraint \rev{only makes sense when it is accumulated over certain periods.} 
   That is to say, \rev{despite tuning the parameters of the controller every day,} we want the cumulative energy consumption to be small during 
   a season or a year\rev{, for example,} rather than on each day. Indeed, it is expected that the system \rev{has to consume more energy} 
   on cold days than on warm ones to optimize \rev{the comfort of the occupants.} 
    
     \textbf{\emph{\rev{Time-varying constraint tracking.} 
    }} Due to changes in energy prices or occupant behaviors, the preference for energy saving or comfort may change. Existing learning-based methods~(e.g., \cite{svetozarevic2022data}) may need to train a controller from scratch again, which can be very \rev{time-consuming} and \rev{data-intensive}. 
    It is desired that the tuning algorithm can respond to such changes of preference rapidly. 

Bayesian optimization~(BO), as a sample-efficient method to optimize black-box functions~\cite{jones1998efficient,xu2022lower} by constructing a stochastic surrogate model, is a promising method to \rev{solve the building controller tuning problem}. 
\rev{However, the vanilla Bayesian optimization method is not readily applicable to address all the aforementioned challenges. In addition, black-box constraints and environmental disturbances need to be incorporated.}

In this paper, we propose to use \rev{our recently proposed} primal-dual contextual Bayesian optimization (PDCBO)~\cite{xu2023primaldual} \rev{method} to automatically and adaptively tune the parameters \rev{of a Proportional Integral (PI) controller, the daytime setpoint temperature, and the pre-heating time} for building thermal control \rev{with time-varying ambient conditions}. \rev{PDCBO has been shown to \rev{asymptotically {achieve the \emph{optimal} performance} while \emph{satisfying constraints} {on average} by adapting to the time-varying environmental changes, which are modeled as contexts} in~\cite{xu2023primaldual}.} Specifically, our contributions \rev{are}: 

\begin{enumerate}
\item The adaptive building controller tuning problem is modeled as an online constrained black-box optimization problem and is then solved with PDCBO method.


\item 
Simulation results show that \rev{PDCBO} can simultaneously optimize the energy consumption and keep the discomfort level below a user-defined upper bound on average in the long term\rev{, contrary to other state-of-the-art constrained BO methods}. 

\item 
When the discomfort threshold is time-varying, 
PDCBO \rev{is again the only method able to rapidly adapt} to the threshold changes and track the constraints well. This highlights \rev{the capability of PDCBO to} automatically explore the allowable discomfort level in order to reduce energy consumption. 

\item 
\rev{When} we instead aim to minimize the discomfort subject to the daily average energy consumption being within a given budget, 
simulation results show that {PDCBO~\cite{xu2023primaldual}} reduces the average discomfort by about \rev{$63\%$} compared to \rev{other safe BO method.} 
\end{enumerate}

\section{Related Work}

Existing research in the multi-objective building optimization literature typically uses a weighted sum of discomfort and energy consumption as the performance target when designing or learning building controllers, e.g. in~\cite{oldewurtel2012use,chen2019gnu,svetozarevic2022data,shen2023bim,yan2022data}. However, with such a formulation, there is no systematic way of choosing the \rev{weighting factor between the two objectives}. Random weights may lead to undesirable consequences, such as intolerable discomfort \rev{for the occupants} or excessive energy usage with very marginal comfort improvement. 

Some research used Pareto optimization~\cite{jin2012facade,diakaki2010multi,chantrelle2011development}, where a set of trade-off optimal solutions --- on the Pareto front --- are derived. However, this approach is typically sample inefficient, since a good approximation of a Pareto front takes a significant number of Pareto optimal points. Furthermore, Pareto optimization typically does not consider the performance during the optimization process and as such cannot be deployed to tune the parameters of real-world \rev{building} systems \rev{online}, where the online performance during tuning also needs to be guaranteed. We refer the reader to~\cite{nguyen2014review} for a detailed review of simulation-based optimization methods applied to building performance analysis. 

\rev{Compared to existing weighted-sum formulations, our online constrained optimization formulation is more desirable due to the following reasons:}

\textbf{\emph{More predictable solutions.}} Different from the weighted sum formulation, which may lead to \rev{undesirable and unpredictable outcomes with a random weight}, the optimal solution of a constrained problem always respects the constraints~(e.g., daily discomfort upper bounded by a threshold) while minimizing the objective~(e.g., energy consumption).  

\textbf{\emph{Easy for preference personalization.}} Different occupants may have different preferences towards comfort and energy saving. Some occupants may want strictly comfortable temperatures regardless of energy consumption. Others may be less sensitive to the temperature and willing to trade some comfort for more energy savings. In a weighted-sum formulation, how to set the weights according to such energy-discomfort preferences is unclear quantitatively, whereas, in a constrained formulation, it can be easily captured by modifying the constraint on discomfort or energy. 

\rev{Energy consumption modeling is a critical problem for building control~\cite{li2014review}. For BO, we use Gaussian process to learn the black-box mapping from control parameters to the closed-loop energy consumption and discomfort due to its high sample efficiency and minimal human effort.} \rev{BO has already been applied to many different energy systems \rev{(including building)} optimization problems, such as 
model calibration~\cite{ZHAN2022112278,CHAKRABARTY2021111460}, chiller plant controller ~\cite{BHATTACHARYA2021111077}, process control~\cite{xu2023config_control}, battery charging~\cite{JIANG2022118244}, and turbine design~\cite{LISICKI20161404}. However, existing applications of Bayesian optimization-based methods either have no constraint consideration at all or can not guarantee long-term constraint feasibility.}

\rev{One of the most popular methods, Constrained Expected Improvement~(CEI)~\cite{gelbart2014bayesian,gardner2014bayesian}, 
encodes the constraint information in the acquisition function. \rev{This method is however not able to guarantee the satisfaction of the desired constraints.} 
To deal with hard safety-critical constraints, Safe BO 
restricts the sampling inside a safe set where feasibility is guaranteed with high probability~\cite{sui2015safe}. More recently, \cite{zhoukernelized} proposed a primal-dual algorithm to solve the constrained Bayesian optimization problem, which has theoretical bounds on the cumulative regret and soft violation. As an extension \rev{to handle contextual inputs}, \textbf{P}rimal \textbf{D}ual \textbf{C}ontextual \textbf{B}ayesian \textbf{O}ptimization~(PDCBO) \rev{was recently developed}~\cite{xu2023primaldual}. 
In this paper, we propose using PDCBO to solve the controller parameter optimization problem. 

} 


%% file: prob_statement.tex
\section{Problem Statement}
\label{sec:prob_state}
We consider \rev{an online} constrained building controller tuning problem \rev{where the parameters are modified each day.} 

\textbf{\emph{Controller structure}} We consider a PI controller to govern the heating of a room, which is currently a common practice in building automation systems. The control law is given as, 
$$u(t)=K_{\mathrm{p}} e(t)+K_{\mathrm{i}} \int_0^t e(\tau) \mathrm{d}\tau,$$
where $K_{\mathrm{p}}$ is the proportional gain, $K_{\mathrm{i}}$ is the integral gain, $e(t)\rev{\coloneqq T_\textrm{setpoint}(t)-T(t)}$ 
is the difference between the temperature setpoint and the measured temperature $\rev{T}$ at time $t$, and $u(t)$ \rev{(after truncated to the interval [0, 1])} is the control input. In our case, the latter represents 
the heating power of the room. 
To incorporate the different thermal requirements during daytime and nighttime \rev{--- e.g., to leverage the fact that the building is vacant at night ---}, we adopt a lower setpoint during the night than at daytime. 

\textbf{\emph{Tuning variables}} Our tuning variables include the proportional and integral gains of the PI controller, and the daytime setpoint temperature. 
To incorporate pre-heating behaviors, we also include the switching time from nighttime mode to daytime mode as a tuning variable. On the other hand, we fix the nighttime setpoint \rev{to $22.5^{\circ}\mathrm{C}$, which is \rev{inside the comfort range of 20\unit{\degree.C} to 23.5\unit{\degree.C} in ASHRAE 55-1992~\cite{taleghani2013review}} and thus a reasonable choice,} \rev{and the switching time from daytime mode to nighttime mode} \rev{at $6$ PM, which is a typical end-of-work time}. 
We use $\theta$ to denote the tuning parameter, which is the concatenation of all tuning variables.

\textbf{\emph{Contextual variables}} We consider the most critical environmental factors, namely the \rev{forecasted} average daily ambient temperature and 
ambient solar irradiation, as contextual inputs of PDCBO. 
We also include the initial temperature at the start of each day~\rev{(midnight)} as an additional contextual variable. 
We use $z_\di$ to denote the \rev{concatenated} contextual variables on the day $\di$.  

\rev{Since the true average ambient temperature and solar irradiation can not be obtained before selecting the new control parameters in reality, we can use the forecast average ambient temperature $T^{\textrm{ambient}}_\di$ and forecast \rev{average} solar irradiation $I_\di$, on each day $\di$.} \rev{The initial room temperature $T^{\textrm{init}}_{\di}$ is also measured}. \rev{The contextual variables are given as $z_\di=(T^{\textrm{ambient}}_\di, I_\di, T^{\textrm{init}}_\di)$.} 

\textbf{\emph{Discomfort quantification}} Similar to what is done in many existing learning-based building control works~(e.g., \cite{svetozarevic2022data}), we define and quantify the discomfort of the occupants as the cumulative temperature deviations from the \rev{potentially time-varying} comfortable temperature range $[T_{\min}(t), T_{\max}(t)]$: 

$$
c_\textrm{discomfort}(T, t)=
\begin{cases}
0, & \textrm{ if }T_{\min}(t)\leq T\leq T_{\max}(t),\\
T-T_{\max}(t), & \textrm{ if } T>T_{\max}(t),\\
T_{\min}(t)-T, & \textrm{ if } T<T_{\min}(t),\\
\end{cases}
$$

We use $D_\di$ to denote the cumulative discomfort on day $\di$, namely the integration of $c_\textrm{discomfort}$ over day $\rev{\di}$: 
\begin{equation}
    D_\di = \int_{\textrm{day }\di}c_\textrm{discomfort}(T(t), t)\mathrm{d} t.~\label{eq:Dndef}
\end{equation}


\rev{Similarly, we compute the energy consumption over a day as:
\begin{equation}
E_\di = \int_{\textrm{day }\di}p(t)\mathrm{d} t,~\label{eq:Endef}
\end{equation}
where $p(t)$ is the power at time $t$, which is proportional to $u(t)$.
}
\rev{Since both $E_\di$ and $D_\di$ are unknown functions of $\theta$ and $z_\di$ corrupted by noise, we introduce Gaussian processes to model them as follows:}
\bse
\begin{align}
E_\di=J(\theta_\di, z_\di)+\nu_\di^J, \\
D_\di=g(\theta_\di, z_\di)+\nu_\di^g,
\end{align}
\label{eqn:GP_modelling}
\ese
where $J$ is assumed to be sampled from the Gaussian process $\mathcal{GP}(\mu_J, k_J(\cdot, \cdot))$ with prior mean $\mu_J$ and prior covariance $k_J(\cdot, \cdot)$ \rev{and $g$ is constructed similarly.} 
Furthermore, $\nu_\di^J\sim\mathcal{N}(0, \sigma_J^2)$ and $\nu_\di^g\sim\mathcal{N}(0, \sigma_g^2)$ are independent identically distributed Gaussian noises.

\textbf{\emph{\rev{Constrained optimization problem.}}} \rev{On the day $\di$, we aim to adaptively select a set of parameters $\theta_\di$} by solving the following online constrained optimization problem, 
\bse
\begin{align}
\min_{\theta\in\Theta}  \qquad& J(\theta, z_\di),\label{eqn:tune_obj} \\
\text{subject to:} \qquad & g(\theta, z_\di)\leq D^{\mathrm{thr}}, \label{eqn:tune_constraint}
\end{align}
\label{eqn:tune_problem}
\ese
 \rev{where $\Theta$ is the set of candidate control parameters~(e.g., a hyperbox) 
 {and $D^{\mathrm{thr}}$ is the maximum amount of allowed comfort violations}}. \rev{Both the objective and the constraint are unknown black-box functions of the control parameter $\theta$ and the contextual variable $z$, which need to be learned online during a period. Alternatively, we get the energy-constrained discomfort minimization formulation by setting $g$ to be \rev{the} objective and $J$ to be below a threshold in the constraint as shown in Eq.~\ref{eqn:tune_energy_constr_problem}.
\bse
\begin{align}
\min_{\theta\in\Theta}  \qquad& g(\theta, z_\di),\label{eqn:discomfort_obj} \\
\text{subject to:} \qquad & J(\theta, z_\di)\leq E^{\mathrm{thr}}, \label{eqn:energy_constr}
\end{align}
\label{eqn:tune_energy_constr_problem}
\ese
\rev{where $E^{\mathrm{thr}}$ is the energy budget.} 
}
 
 \textbf{\emph{{Performance metric.}}} \rev{Although the problem~\eqref{eqn:tune_problem} is formulated as an online problem with instantaneous objective and constraints, we measure the performance of the algorithm using the whole closed-loop trajectory $\{(\theta_\di, z_\di)\}_{\di=1}^{\tdi}$.} \rev{Overall, the metric is the cumulative objective function during a period $\tdi$, that is
 $
   \sum_{\di=1}^\tdi J(\theta_\di, z_\di), 
$ 
while respecting the constraints $g(\theta_\di, z_\di)\leq D^\mathrm{thr}$ \rev{on average} 
}. \rev{To compare the performance of each algorithm, we introduce two running average metrics, 
\bse
\begin{align}
E_\di^\mathrm{avg}\coloneqq\frac{1}{\di}\sum_{k=1}^\di E_k,\label{eqn:avg_eng} \\
D_\di^\mathrm{avg}\coloneqq\frac{1}{\di}\sum_{k=1}^\di D_k.\label{eqn:avg_discft}
\end{align}
\ese
}

 



%% file: case_study.tex

%% file: methodology.tex
\begin{figure*}[t!]
    \centering
    \includegraphics[width=0.88\textwidth]{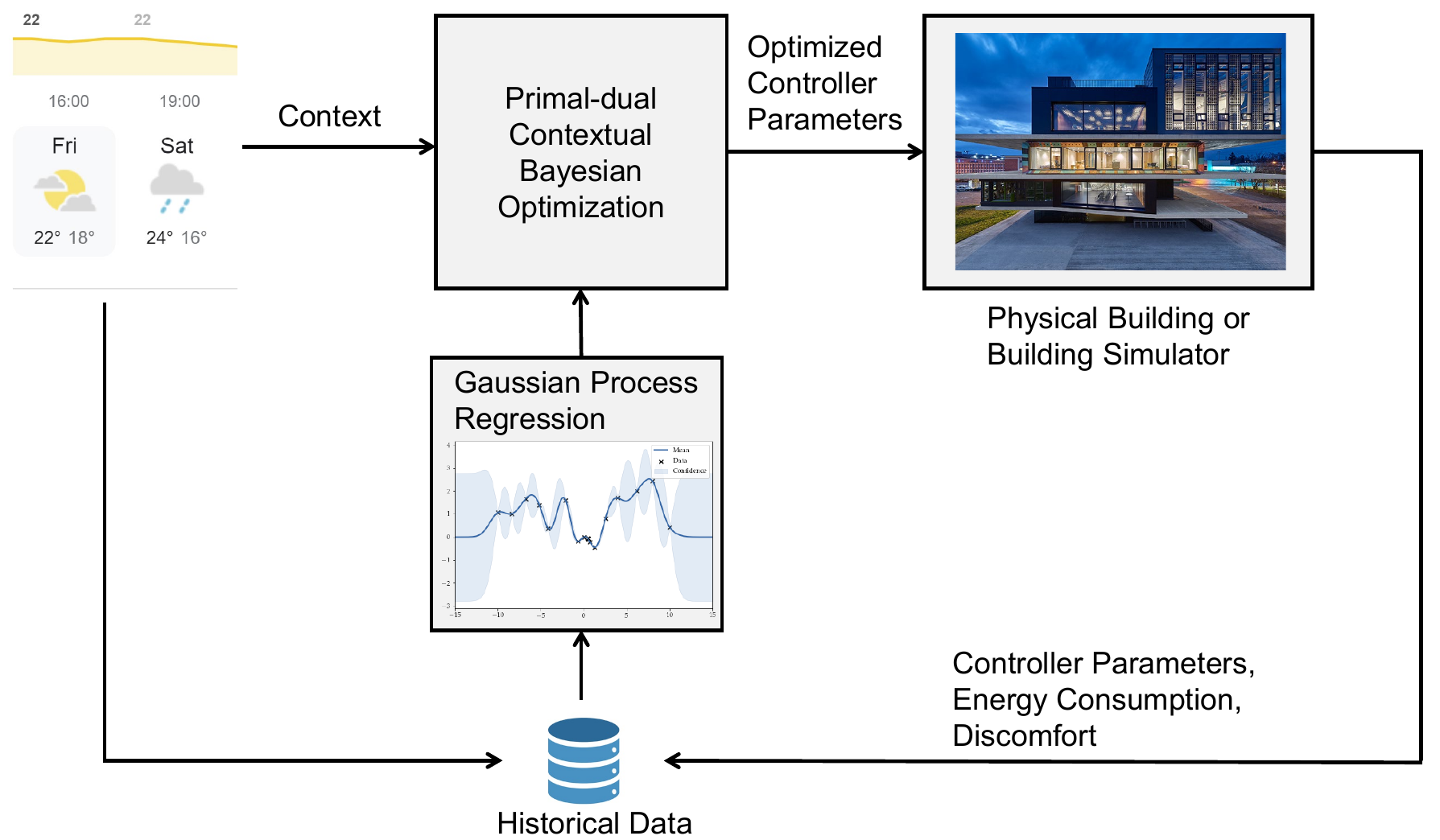}
    \caption{Overview of the proposed controller tuning method. On each day, we get the average temperature and average solar irradiation prediction from given weather forecasts. With the forecasted contexts and the Gaussian process surrogate model, \rev{PDCBO} 
    selects a new set of optimized controller parameters to run the black-box building simulator or conduct a physical experiment. We then collect the new discomfort and energy consumption data and update their GP posteriors. \rev{At the same time, we update the dual variables with the constraints information.} 
    }
    \label{fig:pdcbo}
\end{figure*}
\section{Methodology}
Fig.~\ref{fig:pdcbo} demonstrates an overview of the proposed methodology. On each day $\di$, we collect the contextual information $z_\di$ 
\rev{and use the algorithm detailed in the following sections to choose $\theta_\di$}. 
We then run the simulation 
for the next day with this set of parameters 
\rev{and calculate $E_\di$ and $D_\di$ using the power input and temperature trajectories obtained from simulation or real-world operation.} We then update the Gaussian Processes (GPs) with the new data{, \rev{obtain} the context of the next day, and sample new control parameters}. 

\rev{Note that, given the online setting, this methodology could also directly be applied to a real building, bypassing the need for simulation studies. In this section, we will focus on the Prob.~\eqref{eqn:tune_problem}. The same algorithm can be similarly applied to Prob.~\eqref{eqn:tune_energy_constr_problem}.
}
\subsection{Preliminaries}
\label{subsec:prelim}
\emph{\textbf{Gaussian process regression}}
\rev{We learn $J$ and $g$} using GP regression\rev{, which is a sample-efficient method \rev{to approximate} nonlinear functions}. 
Denoting the set of historical data $$\rev{\mathcal{H}_\di} \coloneqq \rev{\left\{(E_\tau, D_\tau, T^{\textrm{ambient}}_\tau, I_\tau,T^{\textrm{init}}_\tau, \theta_\tau)_{\tau=1}^{\di-1}\right\}}$$ and $x \coloneqq (\theta, z)$, we can derive the posterior distribution of $J$ on day $\di$ as a Gaussian random variable with mean and variance given as~\cite{rasmussen2003gaussian}
\bse
\begin{align}
\hspace{-10mm}\mu_J(x|\mathcal{H}_\di) &= k_J^\top(x, x_{1:\di-1}) (K_J+\sigma_J^2I)^{-1}\Delta y_J + \mu_{J}(x), \\
\hspace{-10mm}\sigma^2_J(x|\mathcal{H}_\di)&=k_J(x, x)\\
&\qquad-k_J^\top(x, x_{1:\di-1})(K_J+\sigma_J^2I)^{-1}k_J(x_{1:\di-1}, x),\nonumber 
\end{align}
\label{eqn:GP_infer}
\ese
where $\Delta y_J=E_{1:\di-1}-\mu_J(x_{1:\di-1})$. 
The posterior Gaussian distribution of $g$ can be similarly derived. \rev{We refer the readers to~\cite{williams2006gaussian} for more details on Gaussian process regression.}

\emph{\textbf{Bayesian optimization (BO)}}
Bayesian optimization is a derivative-free black-box optimization method~(see~\cite{frazier2018tutorial}) which iteratively
\begin{enumerate}
    \item Samples a point to evaluate the black-box function;
    \item Uses \emph{all} the historical samples to construct a surrogate model of the black-box function with GPs.
    \item Finds a new sample that is promising to improve the objective by solving an auxiliary problem constructed with the surrogate model.  
\end{enumerate}

For unconstrained Bayesian optimization, \rev{ a popular way to construct this auxiliary 
problem in step 3 is minimizing 
the \emph{lower confidence bound}}~\cite{srinivas2012information}\footnote{Note that the work~\cite{srinivas2012information} uses the upper confidence bound instead of the lower one since they consider the problem of maximizing $J$ instead of minimizing it.} of the objective, which is defined as
\begin{equation}
   \mu_J(x|\mathcal{H}_n)-\beta_n^{1/2}\sigma_J(x|\mathcal{H}_n), \label{equ:aqu}
\end{equation}
where $\beta_n^{1/2}$ is a positive weighting coefficient that balances exploitation and exploration. The larger $\beta_n$ is, the more likely points with large posterior variance --- i.e., uncertainty --- will be sampled. 





\subsection{PDCBO} \label{sec:PDCBO}
PDCBO is based on the lower confidence bounds \rev{constructed from the posterior distribution of the energy consumption and thermal discomfort}: 
\bse
\begin{align}
 \ubar{J}_n(\theta, z_n)=\mu_J(\theta, z_n|\mathcal{H}_n)-\beta^{1/2}_n\sigma_J(\theta, z_n|\mathcal{H}_n),\\
\ubar{g}_n(\theta, z_n)=\mu_g(\theta, z_n|\mathcal{H}_n)-\beta^{1/2}_n\sigma_g(\theta, z_n|\mathcal{H}_n),
\end{align}
\label{eqn:GP_lcb}
\ese
where \rev{$z_n$ is the observed context at step $n$ and} $\beta^{1/2}_n$ is the positive coefficient 
which balances the tradeoff between \rev{exploitation} and \rev{exploration}, \rev{where exploitation is captured by the posterior mean and the exploration is captured by the posterior standard deviation.} 


\rev{Note that while}~\cite{xu2023primaldual} proposed a rigorous way to select $\beta_\di$ --- \rev{which increases at each time step} --- it can typically be set to be a constant in practice. 

We then relax the constraint in the problem~\eqref{eqn:tune_problem} and write out the corresponding Lagrangian:
\begin{equation}
   \mathcal{L}(\theta, z_\di, \lambda) = J(\theta, z_\di)+\lambda g(\theta, z_\di), 
\end{equation}
\rev{where $\lambda$ is the corresponding dual variable.} 
\rev{Following classical ideas from the constrained optimization literature~\cite{nocedal1999numerical}, we consider the corresponding max-min problem, 
\begin{equation}
   \max_\lambda\min_\theta\mathcal{L}(\theta, z_\di, \lambda). 
\end{equation}
Since both $J$ and $g$ are black-box functions, we \rev{use \eqref{eqn:GP_lcb} to approximate them and} 
solve the inner minimization problem. For the outer maximization problem, on the other hand, we apply a standard dual ascent \rev{step, as detailed in}  
}
Algorithm~\ref{alg:PDCBO}, where $\eta$ is the step size and $\epsilon$ is a small constant shift.  

\begin{algorithm}[!ht]
	\caption{PDCBO for building controller tuning}\label{alg:PDCBO}
	\begin{algorithmic}[1]
	\normalsize
	\State \textbf{Require}: Running horizon $\tdi$, initial dual variable $\lambda_1$ 
	\For{$\di\in[\tdi]$}
	    \State Observe the context $z_\di=(T^{\textrm{ambient}}_\di, I_\di, T^{\textrm{init}}_\di)$
	    \State \textbf{Primal update:} 
	    \vspace{-2ex}
	    $$\theta_\di\in\arg\min\left\{\ubar{J}_\di(\theta, z_\di)+\lambda_\di\ubar{g}_\di(\theta, z_\di)\right\}$$ \label{alg_line:aux_prob} 
	    \vspace{-4ex}
        \State \textbf{Dual update:}
        \vspace{-2ex} 
        $$\lambda_{\di+1}=[\lambda_\di+\eta(\ubar{g}_\di(\theta_\di, z_\di)-D^{\textrm{thr}})+\epsilon]^+$$
        \vspace{-4ex}
        \State \rev{Update the control parameters to $\theta_\di$,} 
        run the closed-loop simulation/experiment, and append the new data $(E_\di, D_\di, T^{\textrm{ambient}}_\di, I_\di, T^{\textrm{init}}_\di, \theta_\di)$ to 
        $\mathcal{H}_\di$.
	    \State {Update the GP posteriors of $J$ and $g$.}
	\EndFor
	\end{algorithmic}
\end{algorithm}


In sharp contrast to existing weighted-sum formulations in model predictive control~\cite{oldewurtel2012use} or reinforcement learning~\cite{svetozarevic2022data} based building control methods, where the weight is static, the weight $\lambda_\di$ in Alg.~\ref{alg:PDCBO} is adaptively 
adjusted each day according to the measurements of the constrained metric. 
Intuitively, if the constraint threshold is violated during one day, the weighting factor will be increased to penalize the constraint more \rev{in the next step}. \rev{Such adaptivity of the weight leads to the automatic balance of energy consumption and discomfort.} 
\rev{In the dual update, we also add a constant drift $\epsilon$} 
to prevent the dual variable from being too small, \rev{which would mean the optimization problem does not consider any constraint anymore and might incur severe violations.}

In this paper, we solve the auxiliary problem in the primal-update step in line~\ref{alg_line:aux_prob} with a grid search algorithm over the set $\Theta$. This is feasible since we are dealing with low dimensions --- $\theta\in\mathbb{R}^4$ ---, \rev{and evaluating both lower confidence bound functions is computationally inexpensive}. 



\subsection{Theoretical Guarantees}

We now restate theoretical guarantees for the performance of Alg.~\ref{alg:PDCBO}, as shown in~\cite{xu2023primaldual}. 

\begin{theorem}[Theorem~6.1,~\cite{xu2023primaldual}]
For kernels with sublinear maximum information gain term~(\rev{see~\cite{srinivas2012information}}) and sufficiently large $\tdi$, by properly setting $\eta$ and $\epsilon$ as shown in~\cite{xu2023primaldual}, we have,
\begin{equation}    
 \sum_{\di=1}^\tdi\left(J(\theta_\di, z_\di)-J(\theta^*(z_\di), z_\di)\right)=\mathcal{O}\left((\gamma^J_\tdi+\gamma^g_\tdi)\sqrt{\tdi}\right),
\end{equation}
where $\theta^*(z_\di)$ is the optimal solution of the Prob.~\eqref{eqn:tune_problem} with contextual variable $z_\di$, $\gamma^J_\tdi$~($\gamma^g_\tdi$, resp.) is the maximum \rev{information gain}, \rev{which measures the complexity of the unknown black-box function $J$~($g$, resp.)~\cite{srinivas2012information}}.
And,
\begin{equation} 
\frac{\sum_{\di=1}^\tdi g(\theta_\di, z_\di)}{\tdi}\leq D^\mathrm{thr}.
\label{eqn:theo_constr_avg}
\end{equation}
\label{thm:main_thm}
\end{theorem}
\rev{Theorem~\ref{thm:main_thm} guarantees that the cumulative gap between the parameters selected by our algorithm and the ground truth time-varying optimal parameters is upper bounded by $\mathcal{O}((\gamma^J_\tdi+\gamma^g_\tdi)\sqrt{\tdi})$. 
Furthermore, the threshold constraint on $g$ is satisfied on average as shown in~\eqref{eqn:theo_constr_avg}. \rev{We refer the interested reader to the theoretical paper~\cite{xu2023primaldual} for more details.} 
}
Specifically, for the commonly used squared exponential~(SE) prior covariance functions, the \rev{cumulative suboptimality} bound reduces to $\mathcal{O}((\log\tdi)^{d+1}\sqrt{\tdi})$~\cite{srinivas2012information}, which is sublinear. \rev{Therefore, the average suboptimality converges to zero, i.e., PDCBO achieves asymptotic optimality on average}. 

\rev{Specific to the building application considered in this paper, Theorem~\ref{thm:main_thm} highlights that PDCBO can asymptotically {achieve the \emph{optimal} building control performance} while \emph{satisfying constraints} {on average}.} 







%% file: result.tex
\section{\rev{Implementations and} Results}
\rev{In this section, we present \rev{some implementation details and then provide} simulation results\rev{, comparing the performance of PDCBO, state-of-the-art constrained BO methods, and fixed control parameters.} We consider both the discomfort-constrained energy minimization problem and the energy-constrained discomfort minimization problem.} 

\begin{figure}[htbp]
    \centering
    \includegraphics[width=0.9\columnwidth]{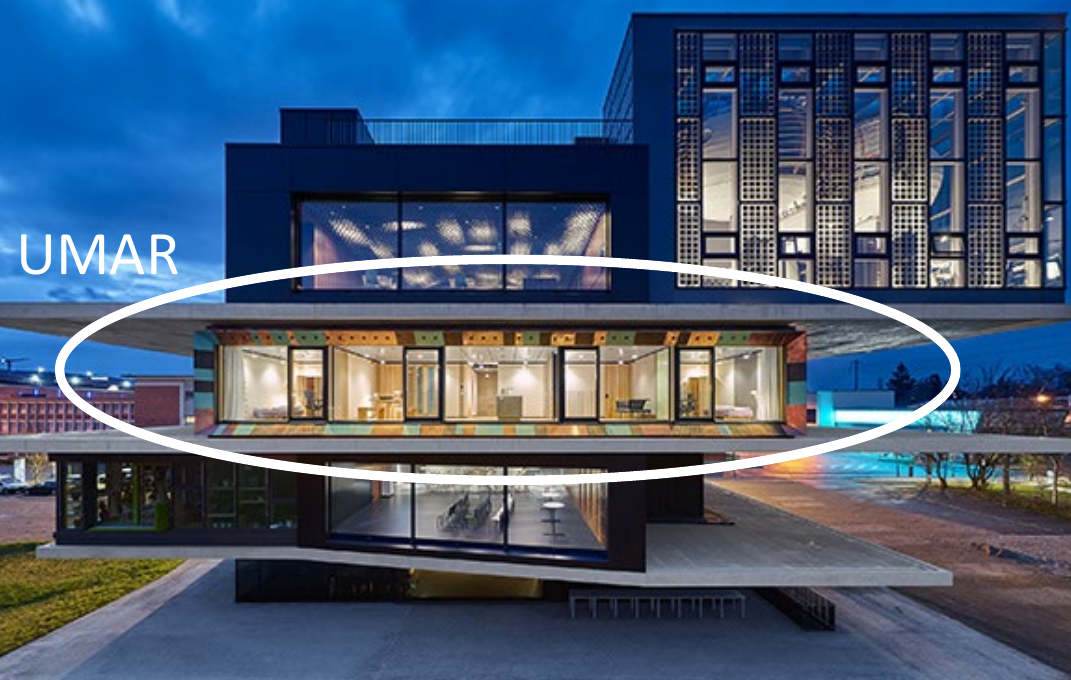}
    \caption{NEST building, Duebendorf, with the UMAR unit in white circle. {\copyright} Zooey Braun, Stuttgart.}
    \label{fig:umar}
\end{figure}


\subsection{\rev{Implementation details}}
\label{sec:imple_det}
\textbf{\emph{Building simulator}} \rev{In this work, we consider the control of the temperature of a single room 
\rev{located in} 
the `Urban Mining and Recycling' (UMAR) unit, pictured in Fig.~\ref{fig:umar}. It is equipped with radiant heating/cooling panels, where hot/cold water is circulated when a valve is opened. We use a digital twin of the room, based on a Physically Consistent Neural Network~(PCNN)~\cite{di2022physically}, as our simulator for evaluation. \rev{Despite relying on Neural Networks (NNs), PCNNs respect the underlying physical laws \textit{by design} while simultaneously achieving state-of-the-art modeling accuracy. In our case, the physical consistency ensures that heating a room increases its temperature, as expected, in contrast to what might happen with classical NNs~\cite{{di2022towards}}. This ensures our algorithm does not pick up spurious behaviors due to modeling inconsistencies. 
On the other hand, PCNNs attain superior prediction accuracy compared to classical physically consistent models~\cite{di2022physically, di2022towards}}.} 




\textbf{\emph{State-of-the-art baselines}} 
We compare our algorithm with two other state-of-the-art Bayesian optimization algorithms: \textsf{SafeOPT}~\cite{fiducioso2019safe} and 
\textsf{CEI}~\cite{gelbart2014bayesian}. \rev{Both baseline algorithms also use Gaussian process to learn the mapping from the control parameters to the daily energy/discomfort.} \rev{SafeOPT aims to not violate the constraints during the whole sampling process and thus is very cautious. CEI algorithm encodes the constraint implicitly in the auxiliary problem but does not explicitly consider feasibility. Therefore, CEI algorithm can be very aggressive and may violate the constraints severely.} For fair comparison purposes, we extend both algorithms to the contextual setting. Specifically, \rev{similarly to the procedure described in Sec.~\ref{sec:prob_state} and Sec.~\ref{sec:PDCBO}, we augment the input space of the Gaussian processes with contextual variables for both SafeOPT and CEI}. 

\textbf{\emph{Choice of parameters}} 
All the GPs use a squared exponential kernel, a common choice in BO applications~(e.g., \cite{xu2022vabo}): 
\begin{equation}
   k(\mathbf{x}, \mathbf{y}) = \sigma_\textrm{SE}^2\exp\left\{-\sum_{i=1}^d\left(\frac{x_i-y_i}{l_i}\right)^2\right\}, 
\label{eq:se_kernel}
\end{equation}
where $\sigma_\textrm{SE}^2$ captures the variation of the function to be modelled, \rev{$l_i$ is the lengthscale that captures the variation rate in the $i$-th input variable}, and $d$ is the dimension of the input variable. We selected those hyperparameters as shown in Table~\ref{tab:hyper_params} \rev{by} \rev{doing maximum likelihood estimation~(MLE) over an initial set of building operational data~\cite{williams2006gaussian}.} 
Concerning the parameters specific to PDCBO, we \rev{manually tried different values} 
and set the dual update step size $\eta=1$ and the lower confidence bound weight $\beta^{1/2}_t=3$\rev{, which works well in this application}. \rev{In this case study, $\epsilon$ 
is manually set to be $0$, which does not \rev{significantly} impact the constraint violations for this particular application.} 




\begin{table}[htbp]
    \centering
    \caption{The choice of hyperparameters for squared exponential kernel, where the heating start time is in the unit of minute.  
    }
    \resizebox{\columnwidth}{!}{
    \begin{tabular}{cccccc}
    \hline
     & $\sigma_\textrm{SE}^2$&  \begin{tabular}{@{}c@{}}{Lengthscale in}\\ {P (log-scale)}\end{tabular}  & \begin{tabular}{@{}c@{}}{Lengthscale in}\\ {I (log-scale)}\end{tabular}& \begin{tabular}{@{}c@{}}{Lengthscale in}\\{daytime setpoint}\end{tabular}  & \begin{tabular}{@{}c@{}}{Lengthscale in}\\ {heating start time}\end{tabular}    \\
         \hline \hline 
    $J$ & 56.7 & 5.9 &3.1& 2.7 & 1290.6  \\
    $g$ &546.1 & 6.0  &8.8 & 5.2  & 1188.0\\
    \hline 
    \end{tabular}
   }
    \label{tab:hyper_params}
\end{table}

\textbf{\emph{Generation of contexts}}
The initial temperature \rev{of the room} is an internal state of the simulator. Ambient temperature and solar irradiation are from the real-world weather dataset recorded at the NEST building of Empa in Zurich. \rev{\rev{In this paper, we use perfect forecast information, i.e., the true average daily ambient temperature and solar irradiation of the next day, as contextual inputs.}} 
\rev{When deploying our method online to the real world, however, real-world temperature and solar irradiation forecasts could be used.} 

\rev{\textbf{\emph{Comfort range and energy price}}} \rev{We adopt a much tighter comfort range ($21$\unit{\degree.C} to $24$\unit{\degree.C} during the nighttime and $23$\unit{\degree.C} to $24$\unit{\degree.C} during the daytime) than the recommended comfort ranges~\cite{taleghani2013review}~(e.g., 20\unit{\degree.C} to 23.5\unit{\degree.C} in ASHRAE 55-1992). Therefore, we allow larger cumulative temperature deviation to capture different levels of comfort requirements.} \rev{To incorporate time-varying electricity prices, we penalize the day-time energy consumption twice, 
which approximately matches the real-world electricity tariff data~\cite{ele_tar}.} 

\rev{
\textbf{\emph{Simulation}} \rev{\rev{In our experiments,} $300$ days are simulated to compare the performance of each controller during the heating season.} 
Since the building needs to be cooled \rev{rather than heated} over the summer and due to data missing issue, the days in our simulation may not be consecutive --- we artificially stitch non-consecutive days \rev{together, using the last simulated room temperature of the previous day as initial value for the next one}. 
}

\rev{Note that control parameters need to be continuously adapted due to the time-varying contexts. 
So we use these running average metrics during the tuning process, instead of choosing one fixed solution over the whole running horizon.} 

\rev{The algorithms are implemented in \textsf{python} and we used \textsf{GPy}~\cite{gpy2014} to implement GPs. 
The code and data can be accessed on 
\url{https://gitlab.nccr-automation.ch/wenjie.cuhk/pdcbo_room_temperature_controller_tuning}. 
}


\begin{figure}[h]
    \centering
    \begin{subfigure}{\columnwidth}
    \centering
    \includegraphics{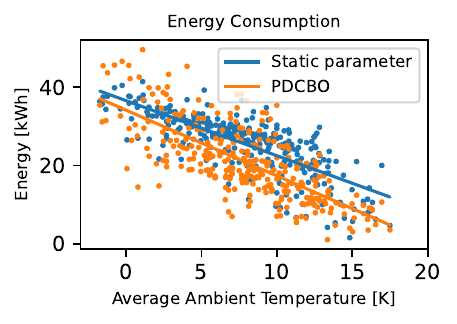}
    \end{subfigure}
    \begin{subfigure}{\columnwidth}
    \centering
    \includegraphics{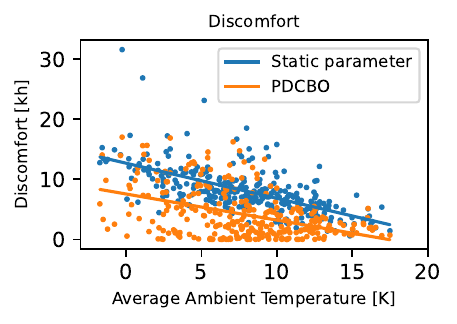}
    \end{subfigure}
    \caption{\rev{Comparison of energy consumption and discomfort with respect to the average ambient temperature. 
    } 
    }
    \label{fig:compare_pdcbo_sample}
\end{figure}

\begin{figure}[h]
    \centering
    \begin{subfigure}{\columnwidth}
    \centering
    \includegraphics{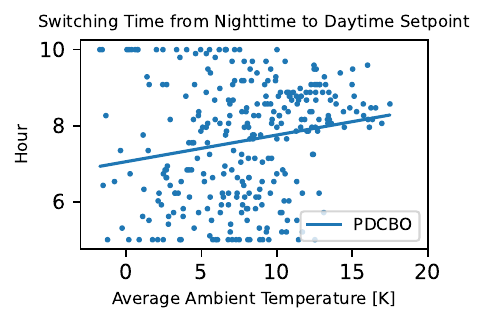}
    \end{subfigure}
    \begin{subfigure}{\columnwidth}
    \centering
    \includegraphics{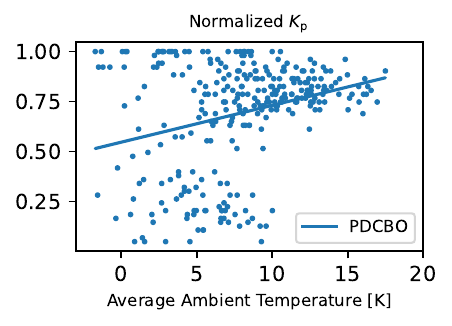}
    \end{subfigure}
    \begin{subfigure}{\columnwidth}
    \centering
    \includegraphics{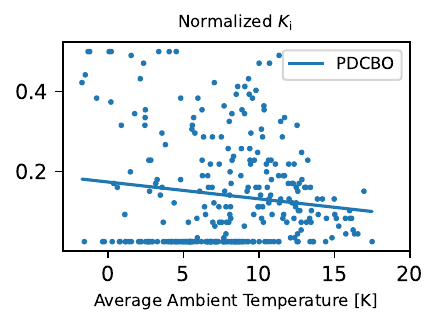}
    \end{subfigure}
    
    \caption{\rev{PDCBO samples of switching time, proportional gain $K_\mathrm{p}$, and integral gain $K_\mathrm{i}$ with respect to ambient average temperature. Both the proportional gain and integral gain are normalized against a base value so that they are unitless.} 
    }
    \label{fig:pdcbo_params_sample}
\end{figure}

\begin{figure*}[h!]
    \centering
    \includegraphics{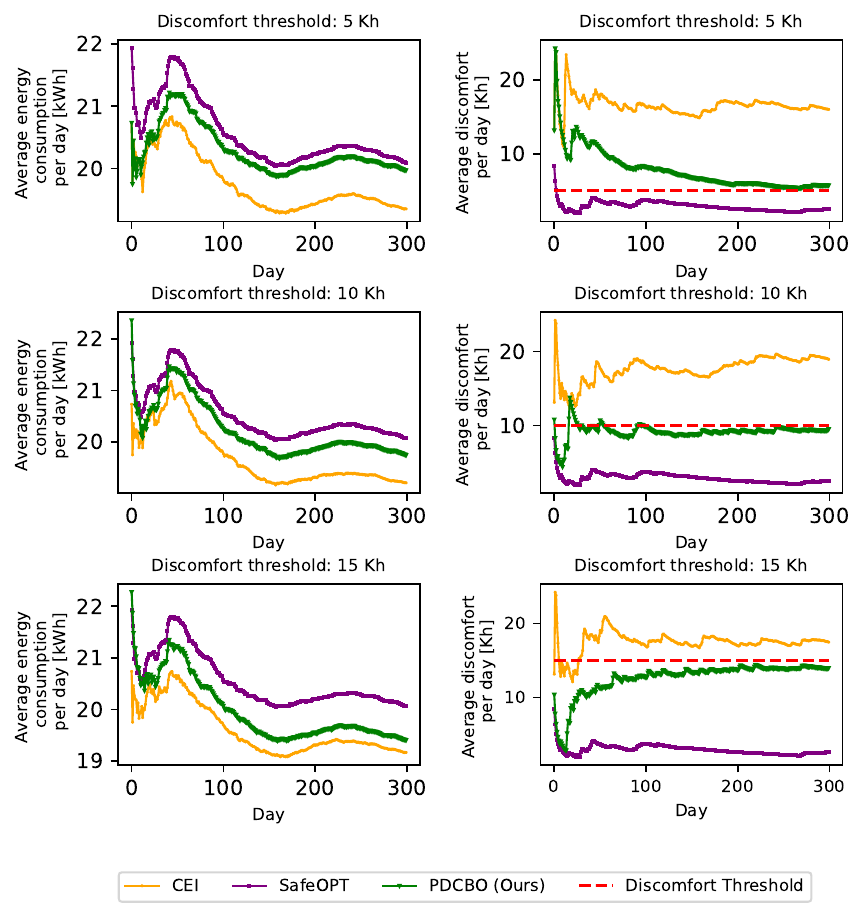}
     \caption{Average energy \rev{consumption} \rev{$E^\mathrm{avg}_\di$} and discomfort \rev{$D^\mathrm{avg}_\di$} per day with discomfort threshold as $5 {\Kh}$, $10 {\Kh}$ and $15 {\Kh}$ per day. 
     \rev{Note that the day-time energy consumption is doubled to reflect the real-world energy tariff.}
     }
    \label{fig:multi_energy_discomfort_with_thr}
\end{figure*}

\begin{figure}[h!]
    \includegraphics{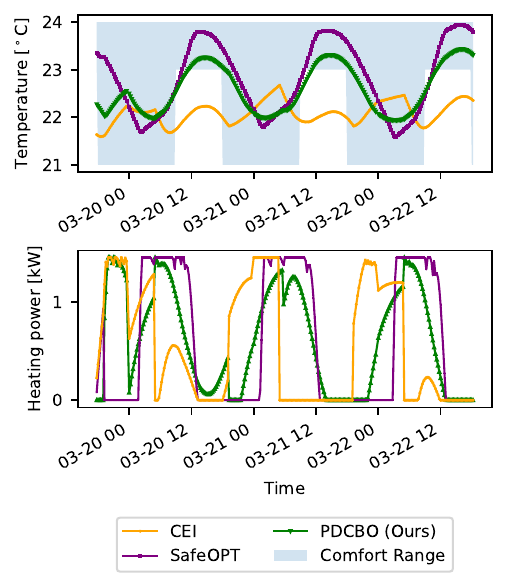}
    \caption{Three-day sample trajectories of the room temperature and heating power. Here, CEI chooses some parameters that simply stops heating during the daytime to save energy, which causes severe discomfort. 
    }
    \label{fig:tmp_vr_traj}
\end{figure}

\subsection{Discomfort constrained energy minimization}
\label{sec:discft_constr_opt}

\subsubsection{Adaptivity Gain Compared to Static Parameters}
\rev{To demonstrate how adaptivity of control parameters benefits the energy saving and discomfort reduction, Fig.~\ref{fig:compare_pdcbo_sample} shows samples of the daily energy consumption and thermal discomfort of the PDCBO algorithm with discomfort threshold $9$ \unit{\kelvin\hour}.} 
\rev{To account for the fact that more energy is naturally needed to maintain the inside temperature in a reasonable range when the temperature outside is lower, we plot both metrics against the mean ambient temperature in a day. For comparison purposes, we also plotted samples corresponding to fixed controller parameters that are manually set by experience. For better visualization, we also plot the linear regression results for both sets of samples. It can be seen that PDCBO can reduce both the energy consumption and the discomfort compared to static parameters. Interestingly, in the low ambient temperature regime~\rev{($\leq10$\unit{\degree.C})}, the discomfort reduction is more significant while in the high ambient temperature regime, the energy reduction is more significant. \rev{This matches our intuition that uncareful heating (e.g., not enough heating power) is more likely to lead to severe discomfort on cold days and thus, there is more room for discomfort reduction on cold days. 
Meanwhile, on warm days, overheating can be an issue with static control parameters and thus there is more room for energy reduction. 
}

\rev{To understand how the parameters are chosen by PDCBO, the coefficients $K_\mathrm{p}$ and $K_\mathrm{i}$ and the} 
switching time from nighttime setpoint to daytime setpoint are also plotted against the average ambient temperature in Fig.~\ref{fig:pdcbo_params_sample}. There is a 
trend to switch the setpoint earlier when the ambient temperature is low, which matches our intuition that pre-heating is needed to avoid low temperatures during working hours on cold days. Interestingly, in the low-temperature regime, PDCBO \rev{tends to favor} 
parameters with small P and large I, contrary to the high-temperature regime. 
\rev{Intuitively, low ambient temperatures have a large `cooling' effect on the room and, therefore, a large integral term is needed to accelerate the heating towards the set-point 
to reduce discomfort and also cancel out the severe offset caused by low ambient temperature. On the contrary, in the high ambient temperature regime, the ambient effect is less severe and small integral term can be adopted to avoid the overshoot and save energy. 
Meanwhile, a large proportional term can help reduce the offset issue in the high ambient temperature regime. Remarkably, such an adaptation policy is automatically found by our PDCBO algorithm.
}

}

\subsubsection{Comparison with SafeOPT and CEI}

\begin{figure}[hbtp]
 \centering
 \includegraphics{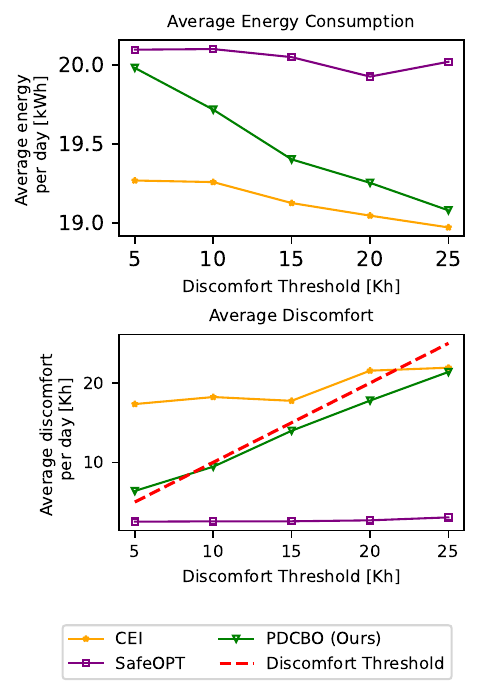}
 \caption{{\rev{Sensitivity analysis to show the impact of different discomfort thresholds on the final average energy consumption ($E_{300}^\mathrm{avg}$ as defined in Eq.~\eqref{eqn:avg_eng}) and the final average discomfort per day ($D_{300}^\mathrm{avg}$ as defined in Eq.~\eqref{eqn:avg_discft}) over 300 days. Note that the daytime energy consumption is doubled here to reflect the real-world electricity price tariff.}} 
 } 
 \label{fig:energy_discomfort_with_thr_diff}
\end{figure}

Fig.~\ref{fig:multi_energy_discomfort_with_thr} 
shows the daily average energy consumption and thermal discomfort obtained by the three methods over 300 days with three different discomfort thresholds $D^{thr}=5,10$, or $15$ $\unit{K.h}$ per day, where $1$ $\unit{K.h}$ represents $1$ Kelvin temperature deviation from comfortable range for $1$ hour. 
As can be seen, \rev{PDCBO} can keep the average discomfort below the user-defined threshold {in the long-term, contrary to CEI, which incurs severe comfort violations. On the other hand, PDCBO can achieve} \rev{significant} energy savings 
compared to {SafeOPT}. Indeed, our method 
uses the allowed maximum discomfort budget and exploits it to achieve more energy savings.   
In sharp contrast, 
SafeOPT is overly cautious and can not achieve all the potential energy savings given the allowed discomfort threshold. 

Fig.~\ref{fig:tmp_vr_traj} gives three-day sample trajectories of the temperature \rev{and the heating power} \rev{after running for 70 days}, \rev{with discomfort threshold at 5 \unit{K.h}}. This confirms that SafeOPT is very cautious and does not explore the energy saving potential of \rev{lowering} the temperature setpoints. On the other hand, CEI samples parameters aggressively\rev{, }leading to large thermal discomfort. As a good compromise, PDCBO identifies 
control parameters that can simultaneously reduce the energy consumption and manage the discomfort well. 

\begin{figure*}[HB]
    \centering
    \begin{subfigure}[b]{\textwidth}
    \centering
    \includegraphics{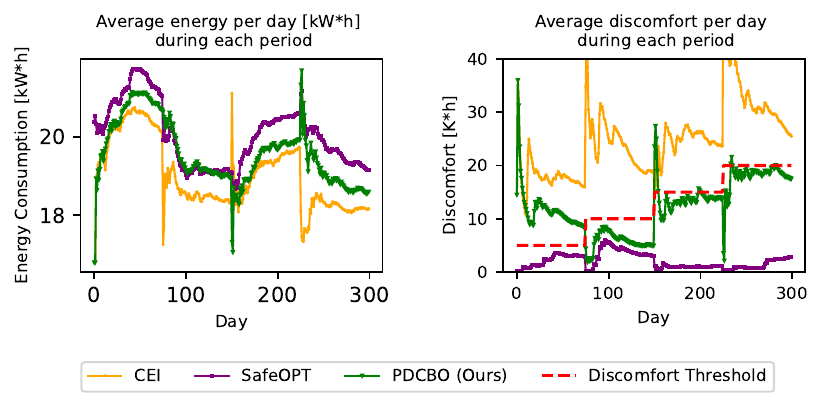}
    \caption{ }
    \label{fig:discomfort_thr_tracking}
    \end{subfigure}
    \hfill 
    \begin{subfigure}[b]{0.3\textwidth}
    \centering
    \includegraphics{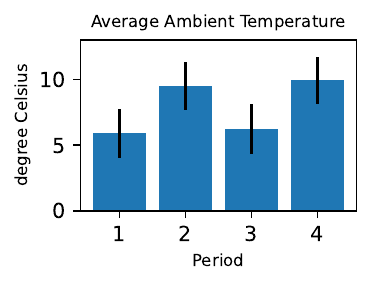}
    \caption{} 
    \label{fig:avg_temp_tracking1}
    \end{subfigure}
    \hfill
    \begin{subfigure}[b]{0.3\textwidth}
    \centering
    \includegraphics{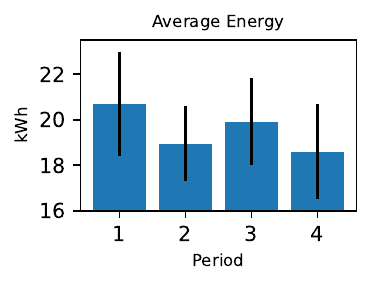}
    \caption{}
    \label{fig:avg_temp_tracking2}
    \end{subfigure}
    \hfill
    \begin{subfigure}[b]{0.3\textwidth}
    \centering
    \includegraphics{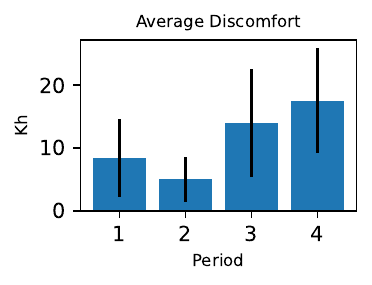}
    \caption{}
    \label{fig:avg_temp_tracking3}
    \end{subfigure}
    \caption{\rev{Discomfort tracking results. (a) Average energy and discomfort per day with time-varying discomfort threshold. The average on day $t$ is taken from day $\tau_0(t)$ to $t$, where $\tau_0(t)$ is the day when the discomfort threshold on day $t$ starts. (b)(c)(d) Average ambient temperature/energy consumption/discomfort during the four periods with different discomfort thresholds. 
    Standard deviations are represented as the black lines.} 
    }
\end{figure*}


\subsubsection{Impact of discomfort threshold} 
To visualize the impact of different thermal comfort thresholds, Fig.~\ref{fig:energy_discomfort_with_thr_diff} presents the average daily energy consumption and comfort violations~\rev{(i.e., the cumulative temperature violations from a comfort range)} obtained by the three methods with a threshold changing from $5$ \unit{K.h} to \rev{$25$ \unit{K.h}}. \rev{Note that the absolute value of the cumulative temperature deviation highly depends on the comfort range adopted. 
As given in Sec.~\ref{sec:imple_det}, the comfort range we adopt is tighter than the standard range. Therefore, we can set the threshold of \rev{discomfort~(quantified by cumulative temperature deviation)} to be from a larger range to capture diversified comfort requirements.} 
We \rev{also} note that \rev{PDCBO} consistently respects the discomfort threshold constraints in all cases in the long-term~(except the minor violation when the discomfort threshold is $5$~\unit{K.h}) while simultaneously leveraging the increasing allowed discomfort budget to reduce the energy consumption. Compared to SafeOPT, PDCBO saves up to \rev{$4.7\%$} energy, as we increase the discomfort threshold 
to \rev{$25$~\unit{K.h}}. Meanwhile, when the discomfort threshold is small, CEI uses less energy than PDCBO but at the cost of severe comfort violations (Tab.~\ref{tab:cei_violations}),
which can be strongly undesired by room occupants. \rev{However, when a large thermal discomfort is tolerable --- when the constraint is easier to meet at $25$~\unit{Kh} ---, CEI can perform slightly better than PDCBO.}

\begin{table}[h]
    \centering 
    \caption{Discomfort violation percentage for CEI method \rev{over the whole running period of $300$ days, which is defined as $\frac{\textrm{Average discomfort}-\textrm{Discomfort threshold}}{\textrm{Discomfort threshold}}$, where the average discomfort is defined as in~Eq.~\eqref{eqn:avg_discft}}. 
    } 
    \begin{tabular}{cccc}
    \hline 
  Discomfort threshold (\unit{\kelvin.h})  & 5    & 10 & 15\\
       \hline  
   Violation Percentage ($\%$)& 247    &  82&  18 \\
         \hline 
    \end{tabular}
    \label{tab:cei_violations}
\end{table}




Remarkably, the average discomfort incurred by PDCBO significantly increases with larger discomfort thresholds $D^\mathrm{thr}$ while the average energy consumption decreases, demonstrating that our method can efficiently exploit the increased allowed thermal discomfort to save more energy. \rev{Such adaptivity to different requirements for comfort is critical since the occupants' preference for thermal comfort~\cite{wang2018individual} and energy consumption can be diverse in practice. On the contrary, the other two baseline methods are either too cautious~(SafeOPT) or overly aggressive~(CEI) and cannot adapt to different discomfort thresholds properly.} 


\subsubsection{Discomfort threshold tracking}

We \rev{now} consider the scenario where the discomfort threshold is time-varying, which \rev{can happen} in practice. For example, when occupants change, new occupants may be willing to trade off more temperature deviations for more energy savings, or vice versa. 
Fig.~\ref{fig:discomfort_thr_tracking} shows the evolution of the average energy consumption and thermal discomfort with a time-varying discomfort threshold. \rev{The discomfort threshold is changed on days 75, 150, and 225. Every time the threshold changes, we reinitialize the calculation of the running average for the discomfort and energy consumption. However, we keep the algorithm continuously running, only modifying the constraint.} For reference, \rev{Fig.~\ref{fig:avg_temp_tracking1} depicts the average ambient temperature during the four periods with different discomfort thresholds. We observe that periods 1 and 3, as well as periods 2 and 4, have similar ambient temperatures. 
Consequently, we focus our analysis on comparing periods 3 and 1 and between periods 4 and 2, respectively.} 

Interestingly, PDCBO closely tracks the change of discomfort threshold (Fig.~\ref{fig:discomfort_thr_tracking}, right) and exploits the more and more allowable temperature deviations to gradually reduce energy consumption, \rev{(comparing period 3 to 1 or period 4 to 2 in Fig.~\ref{fig:avg_temp_tracking2})}. \rev{Note that the energy increase in period 3 (from about day 150 to day 225) as compared to period 2 is due to the decrease of ambient temperature~(see Fig.~\ref{fig:avg_temp_tracking1}).} In contrast and similarly to what was observed in the previous sections, the CEI method is overly aggressive and fails to respect the time-varying discomfort threshold most of the time. SafeOPT, on the contrary, is overly cautious and fails to exploit the time-varying discomfort threshold to reduce the energy consumption. 
\rev{When the discomfort threshold changes, we reset the running averages of energy consumption and discomfort, leading to some spikes in these two metrics.}

\subsection{Energy constrained discomfort minimization}
\label{sec:energy_constr_opt}

\rev{
In this section, we consider the regime where energy consumption is limited to a low budget due to, for example, the energy crisis~\cite{energy_price}, and the pressure of achieving carbon-neutrality targets on time~\cite{carbon_neutrality}\rev{, and we want to maximize the comfort of the occupants with this limited budget.} As compared to Problem~\eqref{eqn:tune_problem}, it is more relevant to bound the average energy consumption for energy constraint~(which directly relates to the electricity bill) during a period.
} 

\begin{figure*}[ht]
    \centering
    \includegraphics{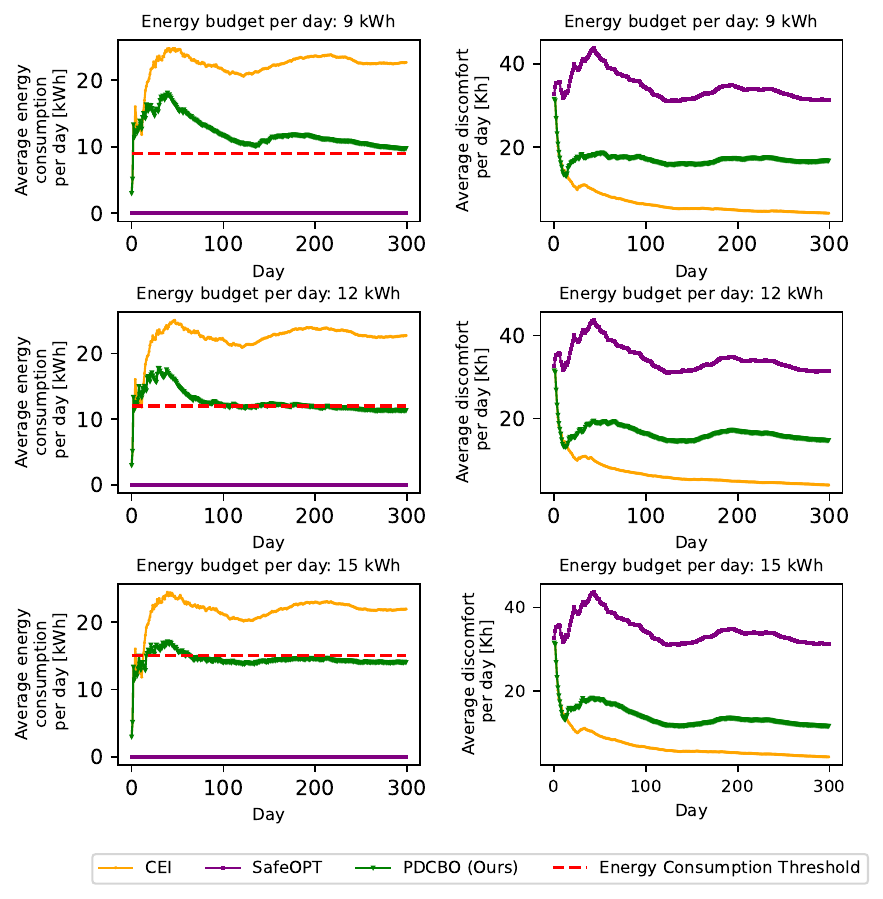}
    \caption{Average energy consumption and discomfort with limited average energy consumption budget as $9 \mathrm{kWh}, 12 \mathrm{kWh}$ and $15 \mathrm{kWh}$ per day shown in dashed red. 
    }
    \label{fig:energy_constrained_e_and_d}
\end{figure*}

\begin{figure}[ht]
    \includegraphics{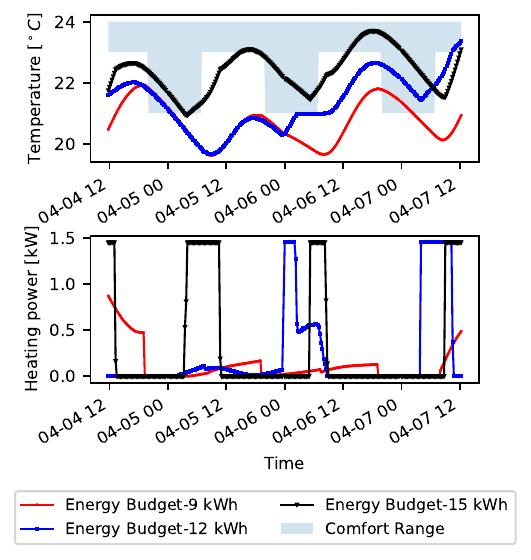}
    \caption{Sample trajectories of room temperature and the corresponding heating power with limited average energy consumption budget as $9 \mathrm{kWh}, 12 \mathrm{kWh}$ and $15 \mathrm{kWh}$ per day. 
    }
    \label{fig:energy_constrained_smp_traj}
\end{figure}

Fig.~\ref{fig:energy_constrained_e_and_d} shows the evolution of \rev{$E^{avg}_\di$ and $D^{avg}_\di$} for the three analyzed methods. It can be observed that PDCBO manages to keep the average energy consumption below the predefined energy consumption threshold in the long term for all three different energy budget constraints. \rev{In sharp contrast, CEI fails to respect the energy budget constraint and can incur an average energy consumption that is more than twice the prescribed energy budget in the long term.} On the other hand, the budgeted energy consumption is well exploited by \rev{PDCBO} to efficiently reduce the discomfort \rev{by respectively $46\%$, $53\%$, and $63\%$, for the three different energy budgets compared to the overly cautious SafeOPT method that tries to keep strictly low energy consumption all the time}. 


\rev{To understand the impact of the energy budget, 
Fig.~\ref{fig:energy_constrained_smp_traj} shows sampled temperature trajectories~(after running for about 200 days) of \rev{PDCBO} with three different 
thresholds. It can be observed that as we gradually increase the budget, the trajectories become better and better and incurs less and less discomfort.} 





%% file: conclusion.tex
\section{Conclusion and discussion}
In this paper, we \rev{{presented a data-driven primal-dual contextual} } Bayesian optimization\rev{~(PDCBO)} \rev{approach to} automatically \rev{tuning} the parameters of a PI controller\rev{, the daytime setpoint of the temperature,} and the pre-heating time to minimize the energy consumption of a room subject to a thermal discomfort constraint\rev{, or minimize the discomfort subject to an energy budget constraint. 
We showcased how PDCBO is able to simultaneously adapt to the time-varying ambient conditions and optimize the objective while satisfying constraints on average.

\rev{When minimizing the energy consumption subject to a thermal discomfort constraint, s}}
imulation results show that \rev{PDCBO} can 
decrease the energy consumption by up to $4.7\%$ compared to state-of-the-art \rev{{\rev{safe} Bayesian optimization}} methods while keeping the average discomfort below the given tolerable threshold. 
\rev{Additionally, 
it can automatically track {time-varying} tolerable thresholds. On the other hand, the other two \rev{constrained} Bayesian optimization methods are either too cautious~(SafeOPT) or overly aggressive~(CEI) and cannot adapt to different discomfort thresholds properly.} 

\rev{
\rev{When we instead optimize the thermal comfort of the occupants subject to a limited energy budget,}} \rev{PDCBO} could reduce the average discomfort \rev{by} up to $63\%$ compared to SafeOPT
, with an average energy consumption below the given budget. \rev{On the other hand, although it achieves lower discomfort, CEI severely violates the energy budget constraint.} 

\rev{
One \rev{promising} direction \rev{to further increase the sample efficiency of BO controller tuning could be to exploit} the grey-box nature of the energy/discomfort functions~(see Eq.~\eqref{eq:Endef} and Eq.~\eqref{eq:Dndef}, which are integrations of a black-box function) to accelerate the tuning of a building controller. Another direction is considering multi-zone building controller tuning, where the layout and geometrical information of the zones may be exploited to improve the tuning efficiency using the recently proposed grey-box BO method~\cite{xu2023bayesian}. } 







%% file: acknowledgements.tex
    \section*{Acknowledgements}
    
This research was supported by the Swiss National Science Foundation under NCCR Automation, grant agreement 51NF40\_180545, and in part by the Swiss Data Science Center, grant agreement C20-13.

    \section*{Declaration of competing interests}
    
The authors declare that they have no known competing financial interests or personal relationships that could have appeared to influence the work reported in this paper.

%% file: main.bbl
\begin{thebibliography}{50}
\expandafter\ifx\csname natexlab\endcsname\relax\def\natexlab#1{#1}\fi
\providecommand{\bibinfo}[2]{#2}
\ifx\xfnm\relax \def\xfnm[#1]{\unskip,\space#1}\fi
\bibitem[{Boodi et~al.(2018)Boodi, Beddiar, Benamour, Amirat, and
  Benbouzid}]{boodi2018intelligent}
\bibinfo{author}{A.~Boodi}, \bibinfo{author}{K.~Beddiar},
  \bibinfo{author}{M.~Benamour}, \bibinfo{author}{Y.~Amirat},
  \bibinfo{author}{M.~Benbouzid},
\newblock \bibinfo{title}{Intelligent systems for building energy and occupant
  comfort optimization: A state of the art review and recommendations},
\newblock \bibinfo{journal}{Energies} \bibinfo{volume}{11}
  (\bibinfo{year}{2018}) \bibinfo{pages}{2604}.
\bibitem[{Oldewurtel et~al.(2012)Oldewurtel, Parisio, Jones, Gyalistras,
  Gwerder, Stauch, Lehmann, and Morari}]{oldewurtel2012use}
\bibinfo{author}{F.~Oldewurtel}, \bibinfo{author}{A.~Parisio},
  \bibinfo{author}{C.~N. Jones}, \bibinfo{author}{D.~Gyalistras},
  \bibinfo{author}{M.~Gwerder}, \bibinfo{author}{V.~Stauch},
  \bibinfo{author}{B.~Lehmann}, \bibinfo{author}{M.~Morari},
\newblock \bibinfo{title}{Use of model predictive control and weather forecasts
  for energy efficient building climate control},
\newblock \bibinfo{journal}{Energy and Buildings} \bibinfo{volume}{45}
  (\bibinfo{year}{2012}) \bibinfo{pages}{15--27}.
\bibitem[{Xiao and You(2023)}]{XIAO2023121165}
\bibinfo{author}{T.~Xiao}, \bibinfo{author}{F.~You},
\newblock \bibinfo{title}{Building thermal modeling and model predictive
  control with physically consistent deep learning for decarbonization and
  energy optimization},
\newblock \bibinfo{journal}{Applied Energy} \bibinfo{volume}{342}
  (\bibinfo{year}{2023}) \bibinfo{pages}{121165}.
\bibitem[{Gao et~al.(2023)Gao, Miyata, and Akashi}]{GAO2023121106}
\bibinfo{author}{Y.~Gao}, \bibinfo{author}{S.~Miyata},
  \bibinfo{author}{Y.~Akashi},
\newblock \bibinfo{title}{Energy saving and indoor temperature control for an
  office building using tube-based robust model predictive control},
\newblock \bibinfo{journal}{Applied Energy} \bibinfo{volume}{341}
  (\bibinfo{year}{2023}) \bibinfo{pages}{121106}.
\bibitem[{Svetozarevic et~al.(2022)Svetozarevic, Baumann, Muntwiler, Di~Natale,
  Zeilinger, and Heer}]{svetozarevic2022data}
\bibinfo{author}{B.~Svetozarevic}, \bibinfo{author}{C.~Baumann},
  \bibinfo{author}{S.~Muntwiler}, \bibinfo{author}{L.~Di~Natale},
  \bibinfo{author}{M.~N. Zeilinger}, \bibinfo{author}{P.~Heer},
\newblock \bibinfo{title}{Data-driven control of room temperature and
  bidirectional {EV} charging using deep reinforcement learning: Simulations
  and experiments},
\newblock \bibinfo{journal}{Applied Energy} \bibinfo{volume}{307}
  (\bibinfo{year}{2022}) \bibinfo{pages}{118127}.
\bibitem[{Lei et~al.(2022)Lei, Zhan, Ono, Peng, Zhang, Hasama, and
  Chong}]{LEI2022119742}
\bibinfo{author}{Y.~Lei}, \bibinfo{author}{S.~Zhan}, \bibinfo{author}{E.~Ono},
  \bibinfo{author}{Y.~Peng}, \bibinfo{author}{Z.~Zhang},
  \bibinfo{author}{T.~Hasama}, \bibinfo{author}{A.~Chong},
\newblock \bibinfo{title}{A practical deep reinforcement learning framework for
  multivariate occupant-centric control in buildings},
\newblock \bibinfo{journal}{Applied Energy} \bibinfo{volume}{324}
  (\bibinfo{year}{2022}) \bibinfo{pages}{119742}.
\bibitem[{Yang et~al.(2015)Yang, Nagy, Goffin, and Schlueter}]{YANG2015577}
\bibinfo{author}{L.~Yang}, \bibinfo{author}{Z.~Nagy},
  \bibinfo{author}{P.~Goffin}, \bibinfo{author}{A.~Schlueter},
\newblock \bibinfo{title}{Reinforcement learning for optimal control of low
  exergy buildings},
\newblock \bibinfo{journal}{Applied Energy} \bibinfo{volume}{156}
  (\bibinfo{year}{2015}) \bibinfo{pages}{577--586}.
\bibitem[{Coraci et~al.(2023)Coraci, Brandi, Hong, and
  Capozzoli}]{CORACI2023120598}
\bibinfo{author}{D.~Coraci}, \bibinfo{author}{S.~Brandi},
  \bibinfo{author}{T.~Hong}, \bibinfo{author}{A.~Capozzoli},
\newblock \bibinfo{title}{Online transfer learning strategy for enhancing the
  scalability and deployment of deep reinforcement learning control in smart
  buildings},
\newblock \bibinfo{journal}{Applied Energy} \bibinfo{volume}{333}
  (\bibinfo{year}{2023}) \bibinfo{pages}{120598}.
\bibitem[{Shen et~al.(2022)Shen, Zhong, Wen, An, Zheng, Li, and
  Zhao}]{SHEN2022118724}
\bibinfo{author}{R.~Shen}, \bibinfo{author}{S.~Zhong},
  \bibinfo{author}{X.~Wen}, \bibinfo{author}{Q.~An},
  \bibinfo{author}{R.~Zheng}, \bibinfo{author}{Y.~Li},
  \bibinfo{author}{J.~Zhao},
\newblock \bibinfo{title}{Multi-agent deep reinforcement learning optimization
  framework for building energy system with renewable energy},
\newblock \bibinfo{journal}{Applied Energy} \bibinfo{volume}{312}
  (\bibinfo{year}{2022}) \bibinfo{pages}{118724}.
\bibitem[{Salsbury(2005)}]{salsbury2005survey}
\bibinfo{author}{T.~I. Salsbury},
\newblock \bibinfo{title}{A survey of control technologies in the building
  automation industry},
\newblock \bibinfo{journal}{IFAC Proceedings Volumes} \bibinfo{volume}{38}
  (\bibinfo{year}{2005}) \bibinfo{pages}{90--100}.
\bibitem[{Stluka et~al.(2018)Stluka, Parthasarathy, Gabel, and
  Samad}]{stluka2018architectures}
\bibinfo{author}{P.~Stluka}, \bibinfo{author}{G.~Parthasarathy},
  \bibinfo{author}{S.~Gabel}, \bibinfo{author}{T.~Samad},
\newblock \bibinfo{title}{Architectures and algorithms for building
  automation—an industry view},
\newblock in: \bibinfo{booktitle}{Intelligent Building Control Systems},
  \bibinfo{publisher}{Springer}, \bibinfo{year}{2018}, pp.
  \bibinfo{pages}{11--43}.
\bibitem[{Fiducioso et~al.(2019)Fiducioso, Curi, Schumacher, Gwerder, and
  Krause}]{fiducioso2019safe}
\bibinfo{author}{M.~Fiducioso}, \bibinfo{author}{S.~Curi},
  \bibinfo{author}{B.~Schumacher}, \bibinfo{author}{M.~Gwerder},
  \bibinfo{author}{A.~Krause},
\newblock \bibinfo{title}{Safe contextual {B}ayesian optimization for
  sustainable room temperature {PID} control tuning},
\newblock in: \bibinfo{booktitle}{Proceedings of the Twenty-Eighth
  International Joint Conference on Artificial Intelligence},
  \bibinfo{organization}{International Joint Conferences on Artificial
  Intelligence}, pp. \bibinfo{pages}{5850--5856}.
\bibitem[{Yan et~al.(2023)Yan, Zhou, and Yang}]{yan2023ai}
\bibinfo{author}{K.~Yan}, \bibinfo{author}{X.~Zhou}, \bibinfo{author}{B.~Yang},
  \bibinfo{title}{{AI} and {I}o{T} applications of smart buildings and smart
  environment design, construction and maintenance}, \bibinfo{year}{2023}.
\bibitem[{Xu et~al.(2022)Xu, Jones, Svetozarevic, Laughman, and
  Chakrabarty}]{xu2022vabo}
\bibinfo{author}{W.~Xu}, \bibinfo{author}{C.~N. Jones},
  \bibinfo{author}{B.~Svetozarevic}, \bibinfo{author}{C.~R. Laughman},
  \bibinfo{author}{A.~Chakrabarty},
\newblock \bibinfo{title}{{VABO}: {V}iolation-{A}ware {B}ayesian {O}ptimization
  for closed-loop control performance optimization with unmodeled constraints},
\newblock in: \bibinfo{booktitle}{2022 American Control Conference (ACC)},
  \bibinfo{organization}{IEEE}, pp. \bibinfo{pages}{5288--5293}.
\bibitem[{Xu et~al.(2023)Xu, Jones, Svetozarevic, Laughman, and
  Chakrabarty}]{xu2023violation}
\bibinfo{author}{W.~Xu}, \bibinfo{author}{C.~N. Jones},
  \bibinfo{author}{B.~Svetozarevic}, \bibinfo{author}{C.~R. Laughman},
  \bibinfo{author}{A.~Chakrabarty},
\newblock \bibinfo{title}{Violation-aware contextual {B}ayesian optimization
  for controller performance optimization with unmodeled constraints},
\newblock \bibinfo{journal}{arXiv preprint arXiv:2301.12099}
  (\bibinfo{year}{2023}).
\bibitem[{Jones et~al.(1998)Jones, Schonlau, and Welch}]{jones1998efficient}
\bibinfo{author}{D.~R. Jones}, \bibinfo{author}{M.~Schonlau},
  \bibinfo{author}{W.~J. Welch},
\newblock \bibinfo{title}{Efficient global optimization of expensive black-box
  functions},
\newblock \bibinfo{journal}{J. Global Optim.} \bibinfo{volume}{13}
  (\bibinfo{year}{1998}) \bibinfo{pages}{455--492}.
\bibitem[{Xu et~al.(2022)Xu, Jiang, Maddalena, and Jones}]{xu2022lower}
\bibinfo{author}{W.~Xu}, \bibinfo{author}{Y.~Jiang}, \bibinfo{author}{E.~T.
  Maddalena}, \bibinfo{author}{C.~N. Jones},
\newblock \bibinfo{title}{Lower bounds on the worst-case complexity of
  efficient global optimization},
\newblock \bibinfo{journal}{arXiv preprint arXiv:2209.09655}
  (\bibinfo{year}{2022}).
\bibitem[{Xu et~al.(2023)Xu, Jiang, Svetozarevic, and Jones}]{xu2023primaldual}
\bibinfo{author}{W.~Xu}, \bibinfo{author}{Y.~Jiang},
  \bibinfo{author}{B.~Svetozarevic}, \bibinfo{author}{C.~N. Jones},
\newblock \bibinfo{title}{Primal-dual contextual {B}ayesian optimization for
  control system online optimization with time-average constraints},
\newblock in: \bibinfo{booktitle}{2023 IEEE 62nd Conference on Decision and
  Control (CDC)}, \bibinfo{organization}{IEEE}.
\bibitem[{Chen et~al.(2019)Chen, Cai, and Berg{\'e}s}]{chen2019gnu}
\bibinfo{author}{B.~Chen}, \bibinfo{author}{Z.~Cai},
  \bibinfo{author}{M.~Berg{\'e}s},
\newblock \bibinfo{title}{Gnu-{RL}: A precocial reinforcement learning solution
  for building {HVAC} control using a differentiable {MPC} policy},
\newblock in: \bibinfo{booktitle}{Proceedings of the 6th ACM International
  Conference on Systems for Energy-Efficient Buildings, Cities, and
  Transportation}, pp. \bibinfo{pages}{316--325}.
\bibitem[{Shen and Pan(2023)}]{shen2023bim}
\bibinfo{author}{Y.~Shen}, \bibinfo{author}{Y.~Pan},
\newblock \bibinfo{title}{{BIM}-supported automatic energy performance analysis
  for green building design using explainable machine learning and
  multi-objective optimization},
\newblock \bibinfo{journal}{Applied Energy} \bibinfo{volume}{333}
  (\bibinfo{year}{2023}) \bibinfo{pages}{120575}.
\bibitem[{Yan et~al.(2022)Yan, Ji, and Yan}]{yan2022data}
\bibinfo{author}{H.~Yan}, \bibinfo{author}{G.~Ji}, \bibinfo{author}{K.~Yan},
\newblock \bibinfo{title}{Data-driven prediction and optimization of
  residential building performance in {S}ingapore considering the impact of
  climate change},
\newblock \bibinfo{journal}{Building and Environment} \bibinfo{volume}{226}
  (\bibinfo{year}{2022}) \bibinfo{pages}{109735}.
\bibitem[{Jin and Overend(2012)}]{jin2012facade}
\bibinfo{author}{Q.~Jin}, \bibinfo{author}{M.~Overend},
\newblock \bibinfo{title}{Facade renovation for a public building based on a
  whole-life value approach},
\newblock in: \bibinfo{booktitle}{Proceedings of Building Simulation and
  Optimisation Conference, Loughborough, UK}, p. \bibinfo{pages}{378e85}.
\bibitem[{Diakaki et~al.(2010)Diakaki, Grigoroudis, Kabelis, Kolokotsa,
  Kalaitzakis, and Stavrakakis}]{diakaki2010multi}
\bibinfo{author}{C.~Diakaki}, \bibinfo{author}{E.~Grigoroudis},
  \bibinfo{author}{N.~Kabelis}, \bibinfo{author}{D.~Kolokotsa},
  \bibinfo{author}{K.~Kalaitzakis}, \bibinfo{author}{G.~Stavrakakis},
\newblock \bibinfo{title}{A multi-objective decision model for the improvement
  of energy efficiency in buildings},
\newblock \bibinfo{journal}{Energy} \bibinfo{volume}{35} (\bibinfo{year}{2010})
  \bibinfo{pages}{5483--5496}.
\bibitem[{Chantrelle et~al.(2011)Chantrelle, Lahmidi, Keilholz, El~Mankibi, and
  Michel}]{chantrelle2011development}
\bibinfo{author}{F.~P. Chantrelle}, \bibinfo{author}{H.~Lahmidi},
  \bibinfo{author}{W.~Keilholz}, \bibinfo{author}{M.~El~Mankibi},
  \bibinfo{author}{P.~Michel},
\newblock \bibinfo{title}{Development of a multicriteria tool for optimizing
  the renovation of buildings},
\newblock \bibinfo{journal}{Applied Energy} \bibinfo{volume}{88}
  (\bibinfo{year}{2011}) \bibinfo{pages}{1386--1394}.
\bibitem[{Nguyen et~al.(2014)Nguyen, Reiter, and Rigo}]{nguyen2014review}
\bibinfo{author}{A.-T. Nguyen}, \bibinfo{author}{S.~Reiter},
  \bibinfo{author}{P.~Rigo},
\newblock \bibinfo{title}{A review on simulation-based optimization methods
  applied to building performance analysis},
\newblock \bibinfo{journal}{Applied Energy} \bibinfo{volume}{113}
  (\bibinfo{year}{2014}) \bibinfo{pages}{1043--1058}.
\bibitem[{Li and Wen(2014)}]{li2014review}
\bibinfo{author}{X.~Li}, \bibinfo{author}{J.~Wen},
\newblock \bibinfo{title}{Review of building energy modeling for control and
  operation},
\newblock \bibinfo{journal}{Renewable and Sustainable Energy Reviews}
  \bibinfo{volume}{37} (\bibinfo{year}{2014}) \bibinfo{pages}{517--537}.
\bibitem[{Zhan et~al.(2022)Zhan, Wichern, Laughman, Chong, and
  Chakrabarty}]{ZHAN2022112278}
\bibinfo{author}{S.~Zhan}, \bibinfo{author}{G.~Wichern},
  \bibinfo{author}{C.~Laughman}, \bibinfo{author}{A.~Chong},
  \bibinfo{author}{A.~Chakrabarty},
\newblock \bibinfo{title}{Calibrating building simulation models using
  multi-source datasets and meta-learned {B}ayesian optimization},
\newblock \bibinfo{journal}{Energy and Buildings} \bibinfo{volume}{270}
  (\bibinfo{year}{2022}) \bibinfo{pages}{112278}.
\bibitem[{Chakrabarty et~al.(2021)Chakrabarty, Maddalena, Qiao, and
  Laughman}]{CHAKRABARTY2021111460}
\bibinfo{author}{A.~Chakrabarty}, \bibinfo{author}{E.~Maddalena},
  \bibinfo{author}{H.~Qiao}, \bibinfo{author}{C.~Laughman},
\newblock \bibinfo{title}{Scalable {B}ayesian optimization for model
  calibration: Case study on coupled building and {HVAC} dynamics},
\newblock \bibinfo{journal}{Energy and Buildings} \bibinfo{volume}{253}
  (\bibinfo{year}{2021}) \bibinfo{pages}{111460}.
\bibitem[{Bhattacharya et~al.(2021)Bhattacharya, Vasisht, Adetola, Huang,
  Sharma, and Vrabie}]{BHATTACHARYA2021111077}
\bibinfo{author}{A.~Bhattacharya}, \bibinfo{author}{S.~Vasisht},
  \bibinfo{author}{V.~Adetola}, \bibinfo{author}{S.~Huang},
  \bibinfo{author}{H.~Sharma}, \bibinfo{author}{D.~L. Vrabie},
\newblock \bibinfo{title}{Control co-design of commercial building chiller
  plant using {B}ayesian optimization},
\newblock \bibinfo{journal}{Energy and Buildings} \bibinfo{volume}{246}
  (\bibinfo{year}{2021}) \bibinfo{pages}{111077}.
\bibitem[{Xu et~al.(2023)Xu, Jiang, Svetozarevic, and
  Jones}]{xu2023config_control}
\bibinfo{author}{W.~Xu}, \bibinfo{author}{Y.~Jiang},
  \bibinfo{author}{B.~Svetozarevic}, \bibinfo{author}{C.~Jones},
\newblock \bibinfo{title}{{CONFIG}: Constrained efficient global optimization
  for closed-loop control system optimization with unmodeled constraints},
\newblock \bibinfo{journal}{IFAC-PapersOnLine}  (\bibinfo{year}{2023}).
\bibitem[{Jiang et~al.(2022)Jiang, Berliner, Lai, Asinger, Zhao, Herring,
  Bazant, and Braatz}]{JIANG2022118244}
\bibinfo{author}{B.~Jiang}, \bibinfo{author}{M.~D. Berliner},
  \bibinfo{author}{K.~Lai}, \bibinfo{author}{P.~A. Asinger},
  \bibinfo{author}{H.~Zhao}, \bibinfo{author}{P.~K. Herring},
  \bibinfo{author}{M.~Z. Bazant}, \bibinfo{author}{R.~D. Braatz},
\newblock \bibinfo{title}{Fast charging design for lithium-ion batteries via
  {B}ayesian optimization},
\newblock \bibinfo{journal}{Applied Energy} \bibinfo{volume}{307}
  (\bibinfo{year}{2022}) \bibinfo{pages}{118244}.
\bibitem[{Lisicki et~al.(2016)Lisicki, Lubitz, and Taylor}]{LISICKI20161404}
\bibinfo{author}{M.~Lisicki}, \bibinfo{author}{W.~Lubitz},
  \bibinfo{author}{G.~W. Taylor},
\newblock \bibinfo{title}{Optimal design and operation of archimedes screw
  turbines using {B}ayesian optimization},
\newblock \bibinfo{journal}{Applied Energy} \bibinfo{volume}{183}
  (\bibinfo{year}{2016}) \bibinfo{pages}{1404--1417}.
\bibitem[{Gelbart et~al.(2014)Gelbart, Snoek, and Adams}]{gelbart2014bayesian}
\bibinfo{author}{M.~A. Gelbart}, \bibinfo{author}{J.~Snoek},
  \bibinfo{author}{R.~P. Adams},
\newblock \bibinfo{title}{Bayesian optimization with unknown constraints},
\newblock in: \bibinfo{booktitle}{Proc. of the 30th Conf. on Uncertainty in
  Artif. Intell.}, UAI'14, \bibinfo{publisher}{AUAI Press},
  \bibinfo{address}{Arlington, Virginia, USA}, \bibinfo{year}{2014}, p.
  \bibinfo{pages}{250–259}.
\bibitem[{Gardner et~al.(2014)Gardner, Kusner, Xu, Weinberger, and
  Cunningham}]{gardner2014bayesian}
\bibinfo{author}{J.~R. Gardner}, \bibinfo{author}{M.~J. Kusner},
  \bibinfo{author}{Z.~E. Xu}, \bibinfo{author}{K.~Q. Weinberger},
  \bibinfo{author}{J.~P. Cunningham},
\newblock \bibinfo{title}{Bayesian optimization with inequality constraints.},
\newblock in: \bibinfo{booktitle}{Proc. of the Int. Conf. on Mach. Learn.},
  volume \bibinfo{volume}{2014}, pp. \bibinfo{pages}{937--945}.
\bibitem[{Sui et~al.(2015)Sui, Gotovos, Burdick, and Krause}]{sui2015safe}
\bibinfo{author}{Y.~Sui}, \bibinfo{author}{A.~Gotovos},
  \bibinfo{author}{J.~Burdick}, \bibinfo{author}{A.~Krause},
\newblock \bibinfo{title}{Safe exploration for optimization with {G}aussian
  processes},
\newblock in: \bibinfo{booktitle}{Proc. of the Int. Conf. on Mach. Learn.}, pp.
  \bibinfo{pages}{997--1005}.
\bibitem[{Zhou and Ji(????)}]{zhoukernelized}
\bibinfo{author}{X.~Zhou}, \bibinfo{author}{B.~Ji},
\newblock \bibinfo{title}{On kernelized multi-armed bandits with constraints},
\newblock in: \bibinfo{booktitle}{Advances in Neural Information Processing
  Systems}.
\bibitem[{Taleghani et~al.(2013)Taleghani, Tenpierik, Kurvers, and Van
  Den~Dobbelsteen}]{taleghani2013review}
\bibinfo{author}{M.~Taleghani}, \bibinfo{author}{M.~Tenpierik},
  \bibinfo{author}{S.~Kurvers}, \bibinfo{author}{A.~Van Den~Dobbelsteen},
\newblock \bibinfo{title}{A review into thermal comfort in buildings},
\newblock \bibinfo{journal}{Renewable and Sustainable Energy Reviews}
  \bibinfo{volume}{26} (\bibinfo{year}{2013}) \bibinfo{pages}{201--215}.
\bibitem[{Rasmussen(2003)}]{rasmussen2003gaussian}
\bibinfo{author}{C.~E. Rasmussen},
\newblock \bibinfo{title}{Gaussian processes in machine learning},
\newblock in: \bibinfo{booktitle}{Summer school on machine learning},
  \bibinfo{organization}{Springer}, pp. \bibinfo{pages}{63--71}.
\bibitem[{Williams and Rasmussen(2006)}]{williams2006gaussian}
\bibinfo{author}{C.~K. Williams}, \bibinfo{author}{C.~E. Rasmussen},
  \bibinfo{title}{Gaussian processes for machine learning},
  volume~\bibinfo{volume}{2}, \bibinfo{publisher}{MIT press Cambridge, MA},
  \bibinfo{year}{2006}.
\bibitem[{Frazier(2018)}]{frazier2018tutorial}
\bibinfo{author}{P.~I. Frazier},
\newblock \bibinfo{title}{A tutorial on {B}ayesian optimization},
\newblock \bibinfo{journal}{arXiv preprint arXiv:1807.02811}
  (\bibinfo{year}{2018}).
\bibitem[{Srinivas et~al.(2012)Srinivas, Krause, Kakade, and
  Seeger}]{srinivas2012information}
\bibinfo{author}{N.~Srinivas}, \bibinfo{author}{A.~Krause},
  \bibinfo{author}{S.~M. Kakade}, \bibinfo{author}{M.~W. Seeger},
\newblock \bibinfo{title}{Information-theoretic regret bounds for {G}aussian
  process optimization in the bandit setting},
\newblock \bibinfo{journal}{IEEE Transactions on Information Theory}
  \bibinfo{volume}{58} (\bibinfo{year}{2012}) \bibinfo{pages}{3250--3265}.
\bibitem[{Nocedal and Wright(1999)}]{nocedal1999numerical}
\bibinfo{author}{J.~Nocedal}, \bibinfo{author}{S.~J. Wright},
  \bibinfo{title}{Numerical optimization}, \bibinfo{publisher}{Springer},
  \bibinfo{year}{1999}.
\bibitem[{Di~Natale et~al.(2022)Di~Natale, Svetozarevic, Heer, and
  Jones}]{di2022physically}
\bibinfo{author}{L.~Di~Natale}, \bibinfo{author}{B.~Svetozarevic},
  \bibinfo{author}{P.~Heer}, \bibinfo{author}{C.~N. Jones},
\newblock \bibinfo{title}{Physically consistent neural networks for building
  thermal modeling: theory and analysis},
\newblock \bibinfo{journal}{Applied Energy} \bibinfo{volume}{325}
  (\bibinfo{year}{2022}) \bibinfo{pages}{119806}.
\bibitem[{Di~Natale et~al.(2023)Di~Natale, Svetozarevic, Heer, and
  Jones}]{di2022towards}
\bibinfo{author}{L.~Di~Natale}, \bibinfo{author}{B.~Svetozarevic},
  \bibinfo{author}{P.~Heer}, \bibinfo{author}{C.~N. Jones},
\newblock \bibinfo{title}{Towards scalable physically consistent neural
  networks: An application to data-driven multi-zone thermal building models},
\newblock \bibinfo{journal}{Applied Energy} \bibinfo{volume}{340}
  (\bibinfo{year}{2023}) \bibinfo{pages}{121071}.
\bibitem[{{EWZ}(2023)}]{ele_tar}
\bibinfo{author}{{EWZ}}, \bibinfo{title}{Current tariffs for the city of
  {Z}urich},
  \bibinfo{howpublished}{\url{https://www.ewz.ch/en/private-customers/electricity/tariffs/overview-of-tariff.html}},
  \bibinfo{year}{since 2023}.
\bibitem[{{GPy}(2012)}]{gpy2014}
\bibinfo{author}{{GPy}}, \bibinfo{title}{{GPy}: A {G}aussian process framework
  in python}, \bibinfo{howpublished}{\url{http://github.com/SheffieldML/GPy}},
  \bibinfo{year}{since 2012}.
\bibitem[{Wang et~al.(2018)Wang, de~Dear, Luo, Lin, He, Ghahramani, and
  Zhu}]{wang2018individual}
\bibinfo{author}{Z.~Wang}, \bibinfo{author}{R.~de~Dear},
  \bibinfo{author}{M.~Luo}, \bibinfo{author}{B.~Lin}, \bibinfo{author}{Y.~He},
  \bibinfo{author}{A.~Ghahramani}, \bibinfo{author}{Y.~Zhu},
\newblock \bibinfo{title}{Individual difference in thermal comfort: A
  literature review},
\newblock \bibinfo{journal}{Building and Environment} \bibinfo{volume}{138}
  (\bibinfo{year}{2018}) \bibinfo{pages}{181--193}.
\bibitem[{ene(2022)}]{energy_price}
\bibinfo{howpublished}{\url{https://www.energypriceindex.com/price-data}},
  \bibinfo{year}{2022}.
\bibitem[{car(2022)}]{carbon_neutrality}
\bibinfo{howpublished}{\url{https://www.un.org/en/climatechange/net-zero-coalition}},
  \bibinfo{year}{2022}.
\bibitem[{Xu et~al.(2023)Xu, Jiang, Svetozarevic, and Jones}]{xu2023bayesian}
\bibinfo{author}{W.~Xu}, \bibinfo{author}{Y.~Jiang},
  \bibinfo{author}{B.~Svetozarevic}, \bibinfo{author}{C.~N. Jones},
\newblock \bibinfo{title}{Bayesian optimization of expensive nested grey-box
  functions},
\newblock \bibinfo{journal}{arXiv preprint arXiv:2306.05150}
  (\bibinfo{year}{2023}).

\end{thebibliography}
